\documentclass[letterpaper,english]{elsart}
\usepackage{times}
\usepackage[T1]{fontenc}
\usepackage[latin1]{inputenc}
\usepackage{prettyref}
\usepackage{float}
\usepackage{amsmath}
\usepackage{graphicx}
\usepackage{amssymb}
\usepackage[authoryear]{natbib}

\makeatletter
\usepackage{epsf}

\usepackage{babel}
\makeatother
\begin{document}
\begin{frontmatter}

\title{Supersonic rupture of rubber }

\author{M. Marder}

\address{Center for Nonlinear Dynamics and Department of Physics\\
The University of Texas at Austin, Austin TX, 78712}

\ead{marder@chaos.ph.utexas.edu}

\begin{abstract}
The rupture of rubber differs from conventional fracture. It is supersonic,
and the speed is determined by strain levels ahead of the tip rather
than total strain energy as for ordinary cracks. Dissipation plays
a very important role in allowing the propagation of ruptures, and
the back edges of ruptures must toughen as they contract, or the rupture
is unstable. This article presents several levels of theoretical description
of this phenomenon: first, a numerical procedure capable of incorporating
large extensions, dynamics, and bond rupture; second, a simple continuum
model that can be solved analytically, and which reproduces several
features of elementary shock physics; and third, an analytically solvable
discrete model that accurately reproduces numerical and experimental
results, and explains the scaling laws that underly this new failure
mode. Predictions for rupture speed compare well with experiment.
\end{abstract}
\end{frontmatter}
\today

\section{Introduction}

\newcommand{\energy}{e}

\newcommand{\lspace}{a}

\newcommand{\MRa}{A}

\newcommand{\MRb}{B}

\newcommand{\widetext}{\relax}

The theory of fracture was originally developed to explain the failure
of brittle materials, where cracks have a number of common features
\citep{Irwin.57,Kanninen.85,Thomson.86}. A stress singularity builds
up in the vicinity of the tip. Stresses diverge as $1/\sqrt{r}$,
where $r$ is the distance to the tip. The displacement of material
near the tip behaves as $\sqrt{r},$ which means that the tip viewed
closely has the shape of a sideways parabola. Energy flows in towards
the crack from far away and concentrates itself at the tip in just
the amount needed to snap bonds and feed other dissipative processes.
Partly for this reason, cracks in tension cannot travel faster than
the Rayleigh wave speed, which is the speed at which sound travels
across a free surface\citep{Broberg.99,Freund.90}.

Experimental evidence has accumulated showing that the rapid rupture
of rubber sheets, such as when one pops a balloon, is different. If
one cuts a horizontal slit in a rubber sheet and stretches it up,
the opening profile is a sideways parabola, as expected for static
cracks. Once the rupture begins to run, however, the characteristic
$\sqrt{r}$ opening displacement disappears, and is replaced by a
wedge-like shape \citep{Deegan.02.rubber}. Furthermore, the rupture
speed exceeds the shear wave speed in advance of the tip \citep{Petersan.04}.
However the extensions in rubber when it ruptures are very large.
The ordinary theory of fracture begins with the assumption that strains
are very small, typically on the order of a fraction of a percent
far ahead of the crack. In rubber, rupture initiates when strains
are on the order of several hundred percent. Therefore it is not clear
how much of fracture mechanics ought to apply to rubber, and whether
the violations of rules about rupture speed are simply the natural
result for a material that breaks for very large extensions, or whether
the mode of failure is something new.

In this article I will explain the case first outlined in \citet{Marder.PRL.05}
that the rupture of rubber is different from conventional fracture.
It is a tensile failure. The ruptures always travel faster than the
shear wave speed. The opening is a wedge that obeys a simple relation
that applies to Mach cones. Stress is singular near the tip, but strain
is not, and material in front of the tip must be brought rather near
the point of failure for ruptures to propagate. 

Buehler, Gao, and Abraham \citeyearpar{Buehler.03} have proposed
that hyperelasticity plays a critical role in dynamic fracture, and
that in particular an increase of sound speed near a crack tip can
allow cracks to travel faster than the distant shear wave speed. In
the analysis of this paper, there is an increase of elastic modulus
near the tip of the rupture, but it is not in the form that Buehler,
Gao, and Abraham proposed. The theory here relies upon an increase
in the modulus of rubber with frequency, rather than upon the increase
in modulus of rubber with extension. The low-frequency modulus of
real rubber certainly increases as rubber stretches towards the breaking
point \cite[p. 2]{Treloar.75}. However, I found in numerical investigations
that the behavior of ruptures does not change appreciably whether
such stiffening prior to breakage is included or not. In the computations
of this paper, Lagrangean sound speeds are either independent of extension
(Sections \ref{sec:Continuum-Neo--Hookean-theory} and \ref{sec:Discrete-Neo-Hookean-theory}),
or else increase as the rubber contracts (Section \ref{sub:Equation-of-Motion}).
In view of the fairly detailed comparison of theory, numerical work,
and experiment obtained in this work, the conclusion is that static
hyperelasticity is not relevant to the supersonic rupture of rubber.

The theory comes in several forms. I begin with a numerical model
of rubber that is fairly realistic and includes most of the physical
features of rubber indicated by experiment. There are only three parameters
in this model not determined directly from experiment, and of those
there is only one to which the fracture dynamics are particularly
sensitive. Next, I note that after stripping some of the realistic
complexity out of the numerical model, the dynamics of ruptures scarcely
change. The resulting theory is so simple that it can be solved analytically.
The analytical solution takes two forms. The first is a continuum
model with a simple failure criterion that leads to compact closed-form
expressions for rupture speed and opening angle. The expressions for
rupture velocity agree with laboratory data within experimental error,
although they disagree with rupture speeds from numerical modeling
by around 10\%. Finally, the discrete rupture theory has a complete
analytical solution. This solution enables one to see the relation
between the conventional theory of fracture and supersonic ruptures,
and is particular how these two types of solution scale in the macroscopic
limit.

The structure of the paper is as follows: Section \ref{sec:Elementary-Theory}
presents some elementary physical ideas to describe rubber rupture.
Section \ref{sec:Continuum-Energy-Functional} lays out the continuum
energy functional for rubber on which much of the subsequent discussion
will be based. Section \ref{sec:Numerical-methodology} develops a
computational method that makes it possible to obtain numerical solutions
that mimic the main features of experimental ruptures. Section \ref{sec:Continuum-Neo--Hookean-theory}
extracts from the numerical model a simple continuum theory and presents
its solution. Section \ref{sec:Discrete-Neo-Hookean-theory} proceeds
further to obtain an exact analytical solution of the full numerical
model, after some simplifications. Section \ref{sec:Comparison-with-experiment}
compares theoretical predictions with the experimental results. There
are also five appendices, which discuss (A) how to obtain an effective
two--dimensional theory from the original three--dimensional theory,
(B) the computation of sound speeds from the two--dimensional continuum
theory, (C) the computation of forces for the numerical model, (D)
a lattice instability found in some of the numerical models, and (E)
the solution of the discrete model by Wiener-Hopf techniques.

\section{Elementary considerations\label{sec:Elementary-Theory}}

Much of this paper will be concerned with the speed of ruptures in
rubber. Therefore it is necessary to begin by describing precisely
how speed will be defined. Suppose that a sound wave or a rupture
travels through a highly stretched rubber sheet. One can choose either
to describe its speed in the laboratory (Eulerian sound speed) or
back in a coordinate system tied to the original location of mass
points (Lagrangean sound speed). To be more explicit, consider a sheet
of rubber lying relaxed in some initial \emph{material reference}
\emph{configuration}. Describe the sheet in this configuration with
the variable $\vec{r,}$ which serves as a label for mass points.
Once the rubber is stretched and begins to move, the new locations
of mass points will be given by the variable $\vec{u.}$ Now consider
some deformation or bump moving through the rubber at constant speed.
The maximum amplitude of the bump is located at some point $\vec{u}(t)$
which changes in time, and the speed at which it moves is the laboratory
velocity. However, there is another velocity, which from a theoretical
point of view is much more natural to employ, and which helps assemble
the experimental results into a compact scaling form. Whenever the
bump is located at $\vec{u,}$ one can identify the original location
$\vec{r}$ of the mass point now at the center of the bump, and $\vec{r(t)}$
also evolves in time. The speed at which $\vec{r(t)}$ travels is
the Lagrangean speed, and unless otherwise specified, sound and rupture
speeds will always be measured in this Lagrangean reference system.
For example, if one considers a bump of small amplitude moving along
$x$ in rubber that has been stretched by a factor of $\lambda_{x}$
above its original length, then the speed of the bump in the laboratory
frame exceeds that in the reference frame by a factor of $\lambda_{x}$
.

\begin{figure}
\begin{center}\includegraphics{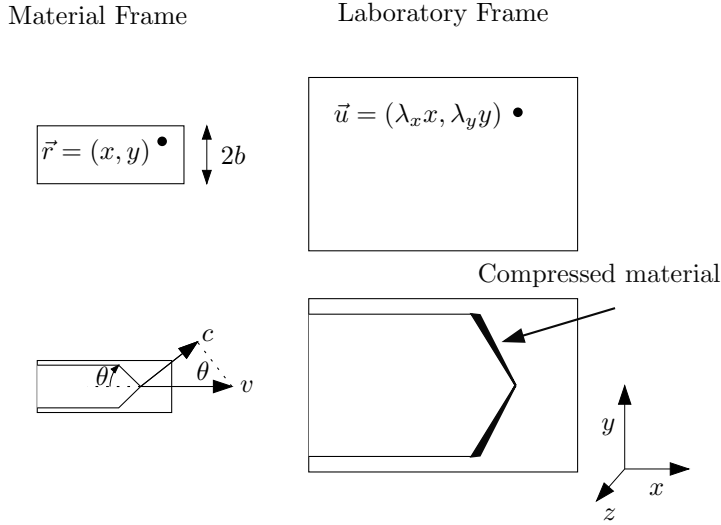}\end{center}

\caption{Illustration of material and laboratory frames used to describe rupture
of rubber. The top left panel shows a rubber sheet before it has been
stretched, and the top right shows it stretched by amounts $\lambda_{x}$
and $\lambda_{y}$ in the $x$ and $y$ directions. The lower right
panel shows a traveling rupture as seen in the laboratory, while the
lower left shows the same rupture back in the material frame. If the
normal velocity of the rupture is $c,$ then the forward velocity
of the rupture must be $v\sin\theta.$ The shock lines shown in the
lower left panel correspond to the leading edge of compressed material
in the lower right panel.\label{cap:Illustration-of-reference}}
\end{figure}

Now consider a thin sheet of rubber that is stretched by factors of
$\lambda_{x}$ and $\lambda_{y}$ in the $x$ and $y$ directions
respectively, as shown in Figure \ref{cap:Illustration-of-reference}.
Stick a pin into the sheet on the left hand side so that a rupture
runs to the right along the $x$ direction. One sees in experiment
\citep{Petersan.04,Deegan.02.rubber} that the rupture consists in
two straight fronts that meet at a point, forming a wedge. Suppose
that the two straight fronts are shock fronts, traveling at the Lagrangean
wave speed $c.$ As is customary for the elementary theory of shocks
\citep[p. 534]{Serway2000}, the speed $v$ of the tip of the rupture
must obey\begin{equation}
\frac{c}{v}=\sin\theta,\end{equation}
 where $\theta$ is the opening angle of the rupture, as shown in
Figure \ref{cap:Illustration-of-reference}. The slope of the upper
face of the shock line is $-\tan\theta,$ which is\begin{equation}
\textrm{material frame slope }=-\frac{1}{\sqrt{v^{2}/c^{2}-1}}.\end{equation}
 In the laboratory, where distances are stretched by factors of $\lambda_{x}$
and $\lambda_{y}$, the slope will intead be\begin{equation}
\textrm{laboratory slope = }-\frac{\lambda_{y}}{\lambda_{x}\sqrt{v^{2}/c^{2}-1}},\label{eq:lab_frame_slope0}\end{equation}
 since if one draws a line of slope $\alpha$ on a sheet of rubber
and stretches it along $x$ by a factor of $\lambda_{x}$, the slope
decreases by a factor of $\lambda_{x}.$ 

Eq. \prettyref{eq:lab_frame_slope0} will re-emerge from detailed
calculations as Eq. \prettyref{eq:lab_face_slope}. The main physical
quantity left undetermined is the rupture velocity $v$. A decent
approximate relation, obtained as Eq. \prettyref{eq:simple_result}
that closes the theory is\[
\lambda_{f}^{2}=\frac{1}{4}\lambda_{x}^{2}+\frac{3}{4}\lambda_{y}^{2}/(1-c^{2}/v^{2}),\]
where $\lambda_{f}$ is an extension at which polymers in rubber snap.
Far ahead of the rupture, the bonds that will eventually be brought
to the snapping point are already stretched an amount $\sqrt{\frac{1}{4}\lambda_{x}^{2}+\frac{3}{4}\lambda_{y}^{2}}$
over their original length. In simple physical terms, then, the assertion
is that just in front of the tip of the rupture, material stretches
by an additional factor of $1/(1-c^{2}/v^{2})$ in the vertical direction.
However, I have not found an elementary argument to produce this relation.

\section{Continuum Energy Functional\label{sec:Continuum-Energy-Functional}}

\subsection{Coordinate system and definition of energy functional\label{sub:Coordinate-system-and}}

Strains in rubber are several hundred percent at rupture and one must
use nonlinear elastic theory to describe the situation \citep{Atkin1980,Ogden1984}.
I state just enough of the theory to establish notation. Adopt a description
of a highly deformed rubber sheet with \begin{equation}
\vec{r}=(r_{x},r_{y})=(x,y)\mapsto\vec{u}=(u^{x},u^{y}).\label{eq:map}\end{equation}
The original location of all mass points is given by $\vec{r}$ and
the location of points after the rubber is moved and stretched is
given by $\vec{u}.$ Note that $\vec{u}$ is measured from the origin,
not from the original location of the mass point $\vec{r}.$ Define
the Lagrangean strain tensor

\begin{equation}
E_{\alpha\beta}\equiv\mbox{$\frac{1}{2}$}\left[\sum_{\gamma}\frac{\partial u^{\gamma}}{\partial r_{\alpha}}\frac{\partial u^{\gamma}}{\partial r_{\beta}}-\delta_{\alpha\beta}\right].\label{eq:Eab}\end{equation}
From this strain tensor one can define three rotationally invariant
quantities. These are\begin{subequations}\begin{eqnarray}
I_{1}^{3D} & = & {\rm Tr\,}E\\
I_{2}^{3D} & = & \sum_{\alpha<\beta}\left[E_{\alpha\alpha}E_{\beta\beta}-E_{\alpha\beta}^{2}\right]\\
I_{3}^{3D} & = & {\rm det}\, E,\end{eqnarray}
\end{subequations}

Rubber is highly incompressible \citep[p. 61]{Treloar.75}. Accordingly,
for a thin sheet of rubber, one can express the thickness at every
point in terms of the strains in the $x-y$ plane, and project the
theory into two dimensions, as discussed in Appendix \ref{sec:Reduction-to-2}.
In two dimensions one has only the two invariants,\begin{equation}
I_{1}={\rm Tr}\, E;\quad I_{2}=E_{xx}E_{yy}-E_{xy}^{2},\label{eq:I1I2}\end{equation}
and using the incompressibility of rubber to solve for $E_{zz}$ one
finds\begin{equation}
E_{zz}=\mbox{$\frac{1}{2}$}\left(\frac{1}{4I_{2}+2I_{1}+1}-1\right),\label{eq:Ezz}\end{equation}
in terms of which the first two of the three--dimensional strain invariants
take the form

\begin{eqnarray}
I_{1}^{3D} & = & I_{1}+E_{zz}\nonumber \\
I_{2}^{3D} & = & I_{2}+E_{zz}I_{1}\label{eq:I3d2d}\end{eqnarray}
The energy density of a thin sheet of rubber is then taken to be given
by some function $\energy$ of $I_{1}$ and $I_{2},$ and the total
energy $U$ is\begin{equation}
U=\rho\int d\vec{r'}\:\energy\left(I_{1}(\vec{r'}),I_{2}(\vec{r'})\right).\label{eq:functional}\end{equation}
The integral is performed in the material frame, and the mass density
$\rho$ is also measured in the material frame.

An energy-conserving equation of motion for this theory is\begin{equation}
0=\frac{\delta}{\delta u^{\gamma}(\vec{r)}}\left[\int d\vec{r}'\ {\textstyle \frac{1}{2}}\rho\left|\dot{\vec{u}}\right|^{2}-\rho\energy\right],\end{equation}
where $\rho$ is the mass per area, again measured in the material
frame. Performing the functional derivatives, one has \begin{equation}
\rho\ddot{u}^{\gamma}=-\frac{\delta U}{\delta u^{\gamma}(\vec{r})}.\label{eq:eq_motion}\end{equation}

Appendix \ref{sec:Reduction-to-2} demonstrates that this equation
of motion for a two-dimensional sheet obtained by calculating $E_{zz}$
through Eq. \prettyref{eq:Ezz} is the same one obtains from the Piola-Kirkhoff
stress tensor after imposing incompressibility and requiring that
the Cauchy stress $T_{zz}$vanish.

\subsection{Sound Speeds}

The experiments by \citet{Petersan.04} that stimulated this study
obtained detailed information about the speed of sound in rubber under
a range of loading conditions. For a while, we found the results puzzling,
but eventually realized that they could all easily be explained by
the Mooney-Rivlin theory. Appendix \ref{sec:Sound-Speeds} contains
a sketch of how to obtain sound speeds from the equation of motion
\prettyref{eq:eq_motion}. Here I record only the final results, all
of which are standard\citep[v. 1, pp. 120, 263]{Eringen.74}. Suppose
that a rubber sheet is strained uniformly with displacement field 

\begin{equation}
\vec{u}=(\lambda_{x}x+s_{xy}y,\lambda_{y}y+s_{yx}x).\label{eq:deformation}\end{equation}
Look for the speed of sound along the $x$ and $y$ axes of a sample
that is extended by the two factors $\lambda_{x}$ and $\lambda_{y}$;
$s_{xy}$ and $s_{yx}$ are included in Eq. \prettyref{eq:deformation}
only because one must be able to perform calculations involving small
virtual shears around this base state. Then there is a longitudinal
sound wave along the $x$ axis whose speed with $s_{xy}=s_{yx}=0$
is

\begin{subequations}

\begin{equation}
c_{xl}^{2}=\frac{\partial^{2}\energy}{\partial\lambda_{x}^{2}}.\label{eq:longitudinal1}\end{equation}
Similarly, the speed of longitudinal waves in the $y$ direction is

\begin{equation}
c_{yl}^{2}=\frac{\partial^{2}\energy}{\partial\lambda_{y}^{2}}.\label{eq:longitudinaly}\end{equation}

There is also a shear wave that travels along $x$ and is polarized
along $y$ with speed

\begin{equation}
c_{xs}^{2}=\frac{\partial^{2}\energy}{\partial s_{yx}^{2}}.\label{eq:shear1}\end{equation}
 Similarly a wave traveling along $y$ and polarized along $x$ has
speed \begin{equation}
c_{ys}^{2}=\frac{\partial^{2}\energy}{\partial s_{xy}^{2}}.\label{eq:shear2}\end{equation}

\label{eq:soundspeeds}\end{subequations}

Alternatively, one can express sounds speeds in terms of derivatives
with respect to strain tensor components. One has for the longitudinal
wave speed along $x$,

\begin{subequations}

\begin{equation}
c_{lx}^{2}=\frac{\partial\energy}{\partial E_{xx}}+\lambda_{x}^{2}\frac{\partial^{2}\energy}{\partial E_{xx}^{2}}\label{eq:longitudinal}\end{equation}

while for the shear wave speed (setting $E_{yx}=E_{xy}$ before taking
the derivatives)\begin{equation}
c_{sx}^{2}=\frac{\partial\energy}{\partial E_{xx}}+\frac{\lambda_{y}^{2}}{4}\frac{\partial^{2}\energy}{\partial E_{xy}^{2}}.\label{eq:shear}\end{equation}
\label{eq:soundspeeds2}\end{subequations}

All of these speeds are Lagrangean speeds, as described in Section
\ref{sub:Coordinate-system-and}. 

Sound speeds provide a convenient way to assemble experimental data
about the constitutive behavior of rubber. In some cases, sound speeds
are measured directly through time of flight, while in other cases
they are measured through small extensions of the sample around a
base state as suggested by Eqs. \prettyref{eq:soundspeeds}. The results
of \citet{Petersan.04} are quite simple. Over a range of biaxial
states where $\lambda_{x}\in[2,3.5]$ and $\lambda_{y}\in[2,3.5]$
the Lagrangean wave speeds appear to be constant, with the longitudinal
wave speed around 20\% greater than the shear wave speed. From Eqs.
\prettyref{eq:soundspeeds2} one finds that this is not possible.
The only way for the longitudinal and shear wave speeds both to be
constant is to adopt $\energy(I_{1},I_{2})\propto I_{1},$ and in
this case the longitudinal and shear wave speeds must be equal.

\begin{figure}
\begin{center}(A)\includegraphics{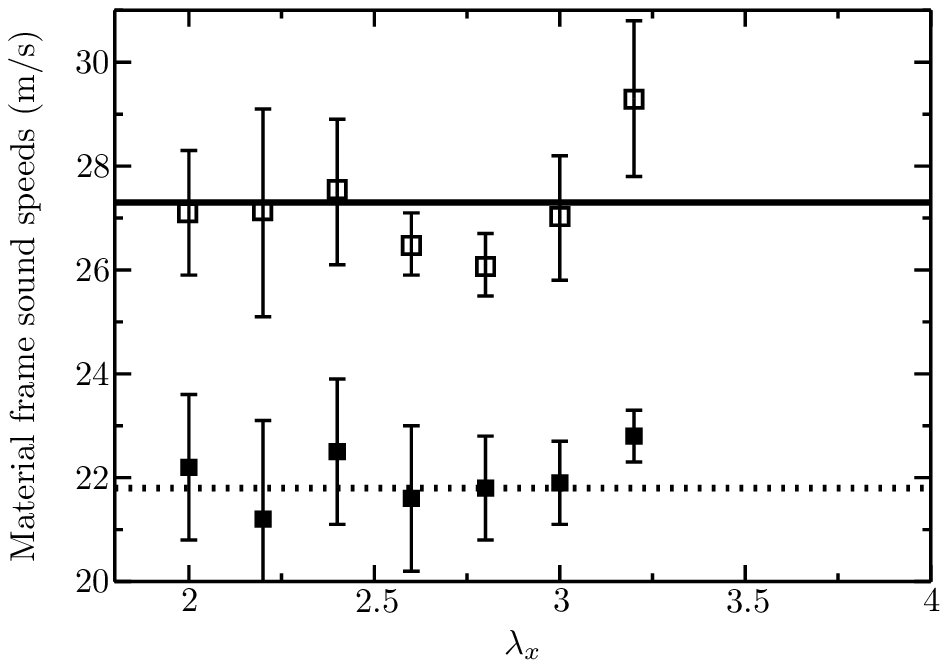}\end{center}

\begin{center}(B)\includegraphics{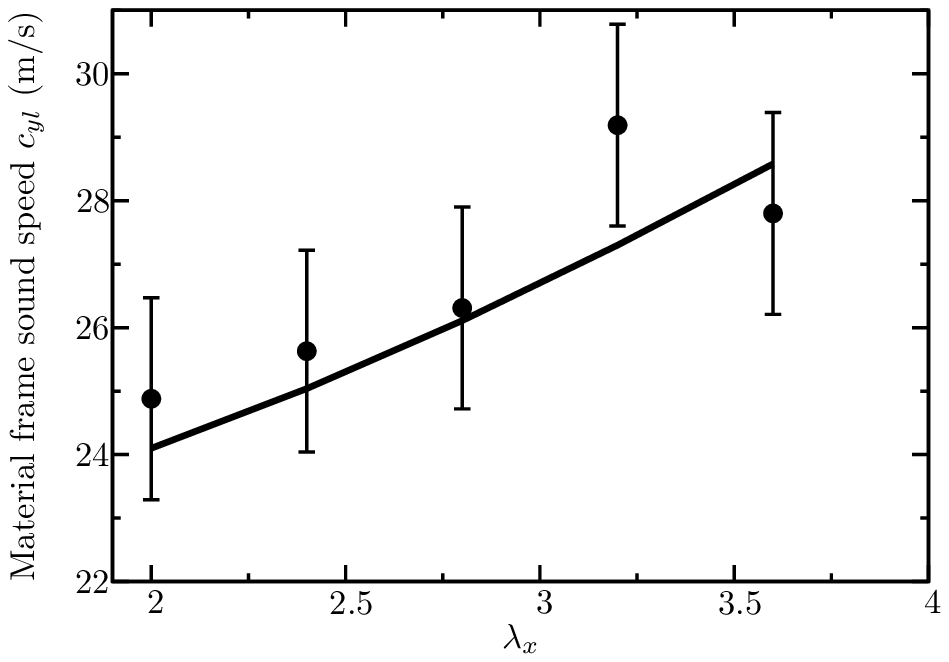}\label{fig:sound_data}\end{center}

\caption{(A) Sound speed data for various values of $\lambda_{x},$ and $\lambda_{y}=3.2.$
Experimental longitudinal ($c_{xl}=\square)$ and shear ($c_{xs}=\blacksquare$)
speeds are shown in the material frame. They are roughly constant,
and can be fit by \prettyref{eq:MR_longitudinal} (---) and \prettyref{eq:MR_shear}
($\dots\dots$), using $\MRa=501\ \mbox{(m/s})^{2},$ $\ \MRb=0.106$.
The shear wave speed comes out to 21.8 m/s and the longitudinal wave
speed for $\lambda_{y}=3.2$ is 27.3. (B) Using the constants $\MRa$
and $\MRb$ obtained from the data in (A), calculate longitudinal
wave speeds in the $y$ direction $c_{yl}$. According to Eq. \ref{eq:MR_soundspeeds},
this is the only speed that varies as a function of $\lambda_{x}$.
The agreement with the data is satisfactory.}
\end{figure}

This apparent difficulty is resolved by examining a bit more carefully
the free energy functional due to Mooney and Rivlin\citep{Treloar.75,Mooney1940,Rivlin1948,Rivlin1948a}.
The Mooney--Rivlin theory says that the free energy density of rubber
is\begin{equation}
U/\rho\equiv\energy=\MRa(I_{1}^{3D}+\MRb I_{2}^{3D}),\end{equation}
where $U$ has units of energy per volume, $\rho$ is mass density,
$\MRa$ is a constant with units of velocity squared, and $\MRb$
is dimensionless. Using Eq. \prettyref{eq:I3d2d} one obtains an effective
two--dimensional Mooney--Rivlin theory\begin{equation}
\energy(I_{1},I_{2})=\MRa\left(I_{1}+\MRb I_{2}+E_{zz}(1+\MRb I_{2})\right)\label{eq:w}\end{equation}
For extensions $\lambda_{x}$ and $\lambda_{y}$ on the order of 2
or greater, $E_{zz}+1/2$ is of order $1/(\lambda_{x}^{2}\lambda_{y}^{2})$
and is at least 64 times smaller than $E_{xx}$ or $E_{yy}$. Therefore,
for the purpose of examining the experiments, it is sufficient to
use\begin{equation}
\energy(I_{1},I_{2})=\MRa(I_{1}+\MRb I_{2})=\MRa\left[(E_{xx}+E_{yy})+\MRb\left(E_{xx}E_{yy}-E_{xy}^{2}\right)\right]\label{eq:w_a}\end{equation}
Employing Eqs. \prettyref{eq:soundspeeds2} and \prettyref{eq:w_a}one
finds for longitudinal and shear wave speeds\begin{subequations}\begin{eqnarray}
c_{xl}^{2} & = & \MRa\left[1+\frac{\MRb}{2}(\lambda_{y}^{2}-1)\right];\ \ \label{eq:MR_longitudinal}\\
c_{yl}^{2} & = & \MRa\left[1+\frac{\MRb}{2}(\lambda_{x}^{2}-1)\right];\label{eq:MR_longitudinal_y}\\
c_{xs}^{2} & = & c_{ys}^{2}=\MRa\left[1-\frac{\MRb}{2}\right].\label{eq:MR_shear}\end{eqnarray}

\label{eq:MR_soundspeeds}\end{subequations}Thus for a Mooney--Rivlin
material the shear wave speed is constant, and the longitudinal speed
along $x$ is independent of $\lambda_{x}$ but depends quadratically
on $\lambda_{y}$. Turning to the experimental data, one finds that
they are consistent with these observations as shown in Fig. \ref{fig:sound_data},
and that one can fix the constants $\MRa$ and $\MRb$. 

For studying the rupture of rubber, the energy density in Eq. \prettyref{eq:w}
is both too simple and too complicated. It is too simple because it
does not account for the fact that when rubber is stretched enough,
the polymers pull apart and the force between adjacent regions drops
irreversibly to zero. It is too complicated because the terms involving
$I_{2}$ and $E_{zz}$ produce nonlinear equations of motion that
are impossible to solve analytically. Therefore, to analyze the problem,
I will pursue two different routes. First, I will discuss numerical
routines that supplement Eq. \prettyref{eq:w} with information about
rupture, toughening, and dissipation, and produce supersonic solutions.
Second, I will isolate from Eq. \prettyref{eq:w} terms that are sufficient
to produce good agreement with numerics and experiment, while simplifying
matters enough to permit analytical solution.

\section{Numerical methodology\label{sec:Numerical-methodology}}

For the problem of fracture, there are great advantages to thinking
in terms of molecular dynamics. From a continuum viewpoint, it is
difficult to understand how to construct a physically sensible theory
where material gives way. From an atomic viewpoint it is easy; when
two atoms are separated by more than a certain distance, they stop
applying force to one another. Therefore, I have found a simple set
of microscopic interactions that produces the Mooney--Rivlin theory
of Eq. \prettyref{eq:w} in the continuum limit. The interacting mass-points
that appear in the theory should not be thought of as atoms. To describe
rubber, they should be thought of as nodes in a cross-linked polymer
network, with a characteristic spacing of around a micron. There are
some possible objections to this approach. Rubber is much more complex
than a triangular lattice, mass is distributed rather than being concentrated
at nodes, and one might worry about the fact that the two-dimensional
array of nodes has been engineered to reproduce dynamics that derive
from projections into two dimensions of three-dimensional equations
of motion. There is no complete answer to these objections; the best
response is to show that the resulting theory provides detailed correspondence
with experiment, and allows much analytical and numerical progress.

Similar numerical techniques have been used before; for example by
\citet{Seung.88}. The techniques of that paper must be extended to
include bond snapping and dissipation, which I carry out here. The
philosophy is also similar to the Virtual Internal Bond method \citep{Gao1998b,Klein1998}
and the Peridynamic Model \citep{Silling.05,Silling.00}, which also
focus on a collection of discrete interacting mass points considerably
larger than atoms in order to obtain rules for fracture. However,
all details of the implementation of this idea are different; the
formulation presented here has the advantage of leading in one case
to a discrete model of nonlinear materials with a complete analytical
solution.

\begin{figure}[h]
\begin{center}\includegraphics{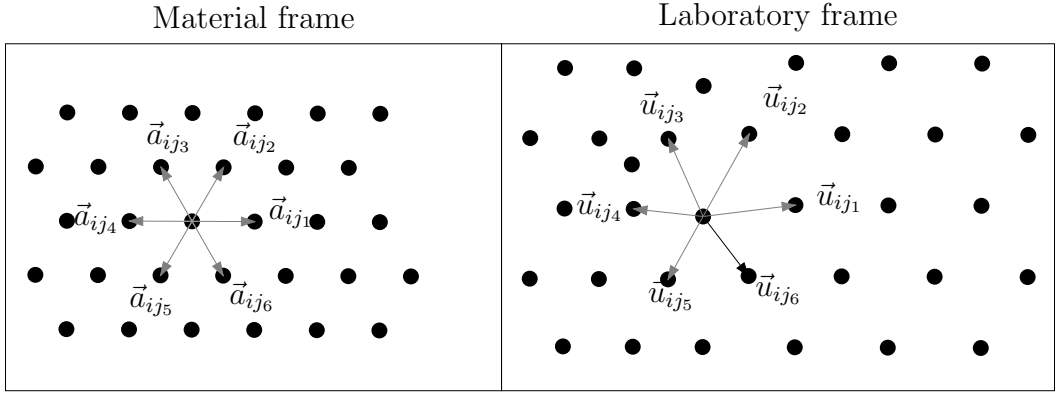}\end{center}

\caption{\label{cap:triang} Diagram showing triangular lattice of lattice
spacing $\lspace$ and nearest-neighbor vectors $\vec{\lspace_{ij}}$
used in this section. The original locations of particles at rest
are given by $\vec{\lspace,}$ while their location in the laboratory
after deformation is given by $\vec{u}.$}
\end{figure}

\subsection{Low-order polynomial terms for microscopic theory}

Consider the triangular lattice depicted in Fig. \ref{cap:triang}.
The figure shows the original locations of all particles prior to
any distortion, denoted by $\vec{\lspace}_{i},$ and the lattice spacing
is $\lspace.$ After distortion, the position of particles in the
laboratory is given by $\vec{u_{i}}$The goal is to construct a theory
for the energy required to displace particles on this lattice that
involves $I_{1}$ and $I_{2}$, and employs quadratic and quartic
functions of displacements $u.$ It is possible to construct such
a theory by considering simple combinations of rotationally invariant
operations on nearest-neighbor vectors. Let $\vec{u_{ij}}\equiv\vec{u_{j}}-\vec{u}_{i},$
let $n(i)$ refer to the nearest neighbors of $i$, and define\begin{subequations}

\begin{equation}
F_{i}=\frac{1}{6}\sum_{j\in n(i)}\begin{cases}
\left(\vec{u}_{ij}\cdot\vec{u}_{ij}-\lspace^{2}\right) & \textrm{if }u_{ij}<\lambda_{f}\\
\lambda_{f}^{2}-\lspace^{2} & \textrm{else}\end{cases}\label{eq:MO1}\end{equation}

\begin{equation}
G_{i}=\frac{1}{9}\sum_{j\in n(i)}\begin{cases}
\left(\vec{u}_{ij}\cdot\vec{u}_{ij}-\lspace^{2}\right)^{2} & \textrm{if }u_{ij}<\lambda_{f}\\
\left(\lambda_{f}^{2}-\lspace^{2}\right)^{2} & \textrm{else}\end{cases}\label{eq:MO2}\end{equation}
\begin{equation}
H_{i}=\frac{1}{27}\sum_{j\neq k\in n(i)}h(u_{ij})h(u_{ik})\left(\vec{u}_{ij}\cdot\vec{u}_{ik}+2\lspace^{2}\right)^{2},\label{eq:MO3}\end{equation}
\begin{equation}
\mbox{and}\quad h(u)=1/(1+e^{(u-\lambda_{f})/u_{s}}).\label{eq:MO4}\end{equation}

\label{eq:MO}\end{subequations}The sums are carried out over the
6 nearest neighbors of point $i$ shown in Fig. \prettyref{cap:triang}.
The terms are constructed so as to become constant and therefore describe
breaking bonds when $u_{ij}$ increases to more than $\lambda_{f}$.
The final term requires a cutoff function $h$ since this is the only
way to ensure both that $H_{i}$ be continuous, and that it settle
down to a constant value when $u_{ij}$ or $u_{ik}$ are large; the
constant $u_{s}$ describes the scale over which $h$ vanishes. Terms
of this form are standard in molecular dynamics (e.g. \citet{Stillinger.85}).
To form a correspondence with continuum theory, suppose that no bonds
are stretched past the breaking point ($u_{ij}<\lambda_{f})$, and
approximate the position of neighbors of point $i$ by

\begin{equation}
u_{ij}^{\alpha}\approx\sum_{\beta}\lspace_{ij}^{\beta}\frac{\partial}{\partial x^{\beta}}u^{\alpha}(\vec{r})\label{eq:MO5}\end{equation}
 Insert Eq. \prettyref{eq:MO5} into Eqs. \prettyref{eq:MO1} to obtain
(using Eq. \prettyref{eq:Eab})

\begin{eqnarray}
F_{i} & \approx & \frac{1}{6}\sum_{j\in n(i)}\left(\sum_{\alpha\beta\gamma}\lspace_{ij}^{\beta}\frac{\partial u^{\alpha}}{\partial x_{\beta}}\frac{\partial u^{\alpha}}{\partial x_{\gamma}}\lspace_{ij}^{\gamma}-\lspace^{2}\right)\label{eq:Fi}\\
 & = & \frac{1}{6}\sum_{j\in(i)}\left(\sum_{\beta\gamma}\lspace_{ij}^{\beta}\left[2E_{\beta\gamma}+\delta_{\beta\gamma}\right]\lspace_{ij}^{\gamma}-\lspace^{2}\right)\nonumber \\
 & = & (E_{xx}+E_{yy})\lspace^{2}=I_{1}\lspace^{2},\end{eqnarray}

\begin{eqnarray}
G_{i} & \approx & \frac{1}{9}\sum_{j\in n(i)}\left(\sum_{\alpha\beta\gamma}\lspace_{ij}^{\beta}\frac{\partial u^{\alpha}}{\partial x_{\beta}}\frac{\partial u^{\alpha}}{\partial x_{\gamma}}\lspace_{ij}^{\gamma}-\lspace^{2}\right)^{2}\nonumber \\
 & =\frac{1}{9} & \sum_{j\in n(i)}\left(\sum_{\alpha\beta\gamma}\lspace_{ij}^{\beta}\left[2E_{\beta\gamma}+\delta_{\beta\gamma}\right]\lspace_{ij}^{\gamma}-\lspace^{2}\right)^{2}\nonumber \\
 & = & \left[(E_{xx}+E_{yy})^{2}+\frac{4}{3}\left(E_{xy}^{2}-E_{xx}E_{yy}\right)\right]\lspace^{4}\nonumber \\
\noalign{\vskip-.1in}\nonumber \\
 & = & \left[I_{1}^{2}-\frac{4}{3}I_{2}\right]\lspace^{4}\label{eq:Gi}\end{eqnarray}

\begin{eqnarray}
H_{i} & \approx & {\textstyle \frac{1}{27}}\sum_{j\neq k\in n(i)}\left(\sum_{\alpha\beta\gamma}\lspace_{ij}^{\beta}\frac{\partial u^{\alpha}}{\partial x_{\beta}}\frac{\partial u^{\alpha}}{\partial x_{\gamma}}\lspace_{ik}^{\gamma}+2\lspace^{2}\right)^{2}\nonumber \\
 & =\frac{1}{27} & \sum_{j\neq k\in n(i)}\left(\sum_{\alpha\beta\gamma}\lspace_{ij}^{\beta}\left[2E_{\beta\gamma}+\delta_{\beta\gamma}\right]\lspace_{ik}^{\gamma}+2\lspace^{2}\right)^{2}\label{eq:Hi}\\
 & = & \left[(E_{xx}+E_{yy})^{2}+\frac{20}{9}\left(E_{xy}^{2}-E_{xx}E_{yy}\right)+4\right]\lspace^{4}\nonumber \\
\noalign{\vskip-.1in}\nonumber \\
 & = & \left[I_{1}^{2}-\frac{20}{9}I_{2}+4\right]\lspace^{4}.\end{eqnarray}

Comparing Eqs. \prettyref{eq:Gi}, \prettyref{eq:Fi}, and \prettyref{eq:Hi}
with Eq. \prettyref{eq:I1I2} gives

\begin{equation}
I_{1}^{i}=\frac{F_{i}}{\lspace^{2}}\label{eq:J1i}\end{equation}

\begin{equation}
I_{2}^{i}=\frac{3}{4}\frac{1}{\lspace^{4}}\left(F_{i}^{2}-G_{i}\right),\label{eq:J2ia}\end{equation}

or alternatively,\begin{equation}
I_{2}^{i}=\frac{9}{8}\frac{1}{\lspace^{4}}\left(G_{i}-H_{i}+4\right).\label{eq:J2ib}\end{equation}

Numerically, Eq. \prettyref{eq:J2ia} is much less costly to compute
than Eq. \prettyref{eq:J2ib}. However, Eq. \prettyref{eq:J2ia} has
the unfortunate property that under biaxial strain, spatially uniform
states are unstable when this representation of $I_{2}$ is employed.
Particles bunch up in a non--uniform way within each unit cell, forming
stripes on a microscopic scale, as shown in Appendix C. It could be
that this behavior is related to the physical phenomenon of strain
crystallization \citep[p. 20]{Treloar.75}. However, as strain crystallization
does not occur experimentally in the range of extensions where ruptures
are observed, I have largely employed Eq. \prettyref{eq:J2ib} in
preference to Eq. \prettyref{eq:J2ia}.

\subsection{Specification of numerical energy functional}

To form a numerical representation of Eq. \prettyref{eq:functional},
take

\begin{equation}
U=m\sum_{i}\energy(I_{1}^{i},I_{2}^{i}),\label{eq:Umic}\end{equation}
where $m$ is the mass in a unit cell. Then if $\Omega$ is the volume
of a unit cell, \begin{equation}
U\approx\frac{m}{\Omega}\int d\vec{r}\ \energy\big(I_{1}(\vec{r}),I_{2}(\vec{r})\big),\end{equation}
so since $m/\Omega=\rho,$ $\energy$ in Eq. \prettyref{eq:Umic}corresponds
to $\energy$ in the continuum theory, and has units of velocity squared.
In particular, for the Mooney--Rivlin theory, one has

\begin{equation}
\energy(I_{1}^{i},I_{2}^{i})=\MRa(I_{1}^{i}+\MRb I_{2}^{i}+E_{zz}^{i}(1+\MRb I_{2}^{i}),\label{eq:Umic_MR}\end{equation}
where $I_{1}^{i}$ is given by Eq. \prettyref{eq:J1i}, $I_{2}^{i}$
is given either by Eqs. \prettyref{eq:J2ia} or \prettyref{eq:J2ib}
, and $E_{zz}^{i}$ is given by\prettyref{eq:Ezz}, with $I_{1}^{i}$
and $I_{2}^{i}$ substituted for $I_{1}$ and $I_{2}$ .

\subsection{Equation of Motion\label{sub:Equation-of-Motion}}

Given the energy functional (\ref{eq:Umic}) one can obtain the force
on every particle and therefore an equation of motion. In addition
to the conservative force resulting from derivatives of the energy,
add Kelvin dissipation, so that the complete equation of motion reads\begin{equation}
m\ddot{u}_{i}^{\alpha}=-\partial U/\partial u_{i}^{\alpha}+\sum_{j\in n(i)}\frac{2m\MRa\beta}{3\lspace^{2}}\dot{u}_{ij}^{\alpha}\theta(\lambda_{f}-u_{ij}).\label{eq:motion}\end{equation}
The final term in Eq. \prettyref{eq:motion} represents the Kelvin
dissipation, and is the simplest dissipative term permitted by symmetry.
The motion of each mass point dissipates some energy in proportion
to its velocity relative to each neighbor. This dissipation vanishes
when the bond between two neighbors breaks. 

The computation of $\partial U/\partial u_{i}^{\alpha}$ is not particularly
difficult as force computations go. A formalism that makes it easy
to exploit symmetry to reduce the amount of computation is briefly
described in Appendix \ref{sec:Force-computation}.

One final rule is employed in the numerical runs, although it is not
indicated explicitly in Eq. \prettyref{eq:motion}. Whenever some
bond $u_{ij}$ drops to a length less than $1.5\lspace$, the failure
extension $\lambda_{f}$ for the remaining bonds attached to nodes
$i$ and $j$ increases. Without some rule of this type, the back
faces of the crack disintegrate. The reason for this rule will be
explained in Section \ref{sub:Disintegration-of-back}.

\begin{figure}
\begin{center}\includegraphics[%
  width=0.40\columnwidth,
  keepaspectratio]{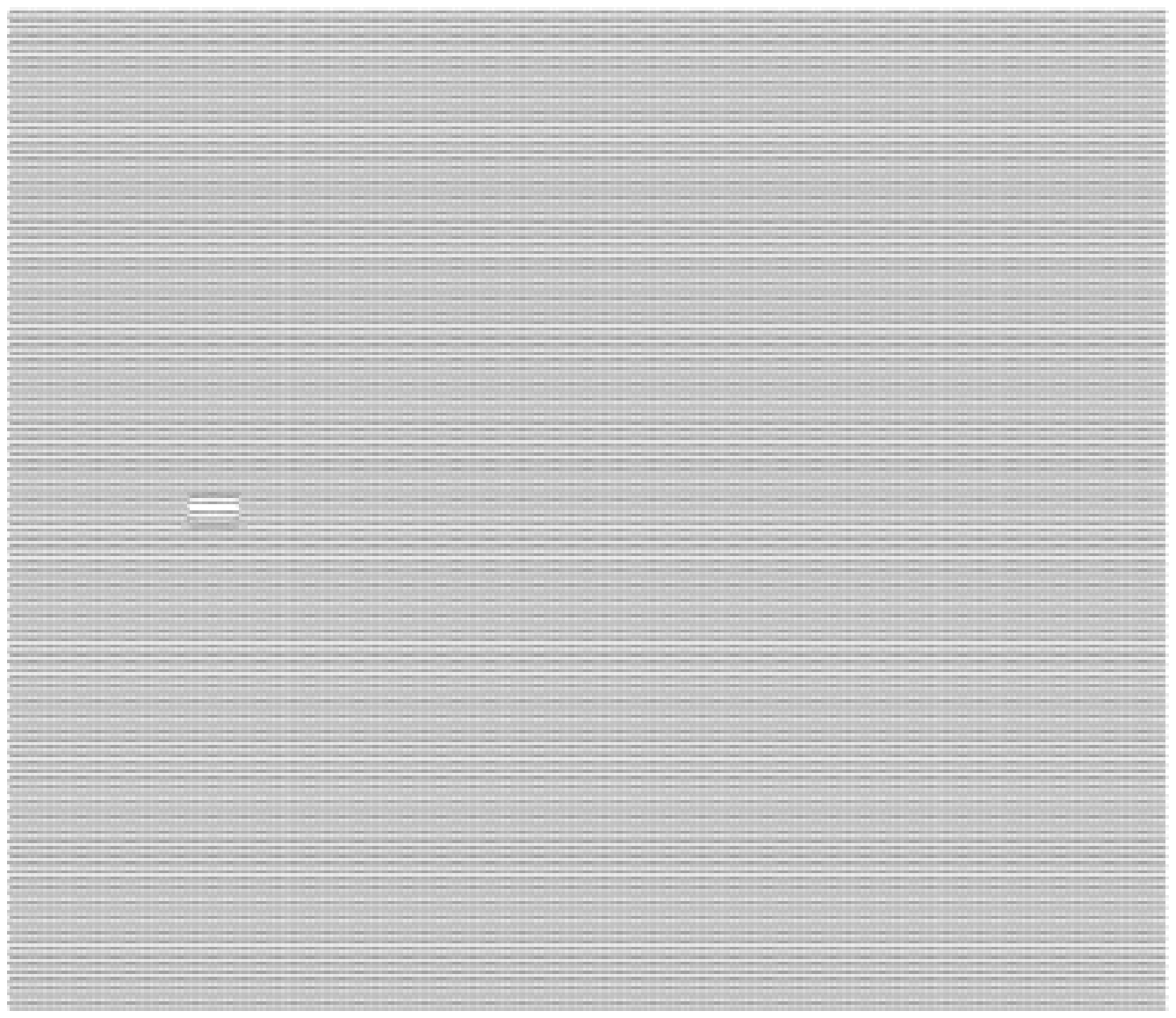}\includegraphics[%
  width=0.40\columnwidth,
  keepaspectratio]{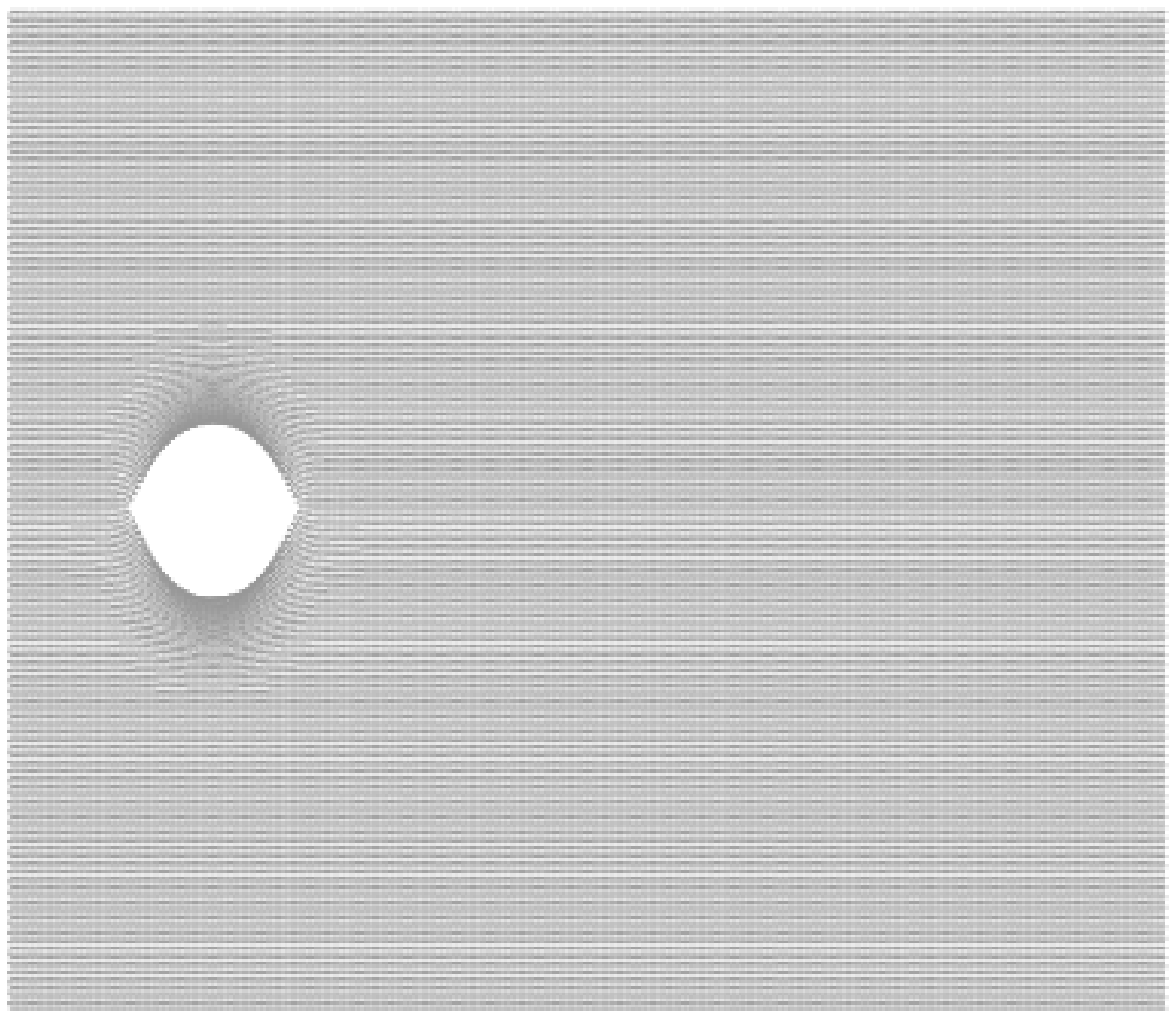}\end{center}

\begin{center}\includegraphics[%
  width=0.40\columnwidth,
  keepaspectratio]{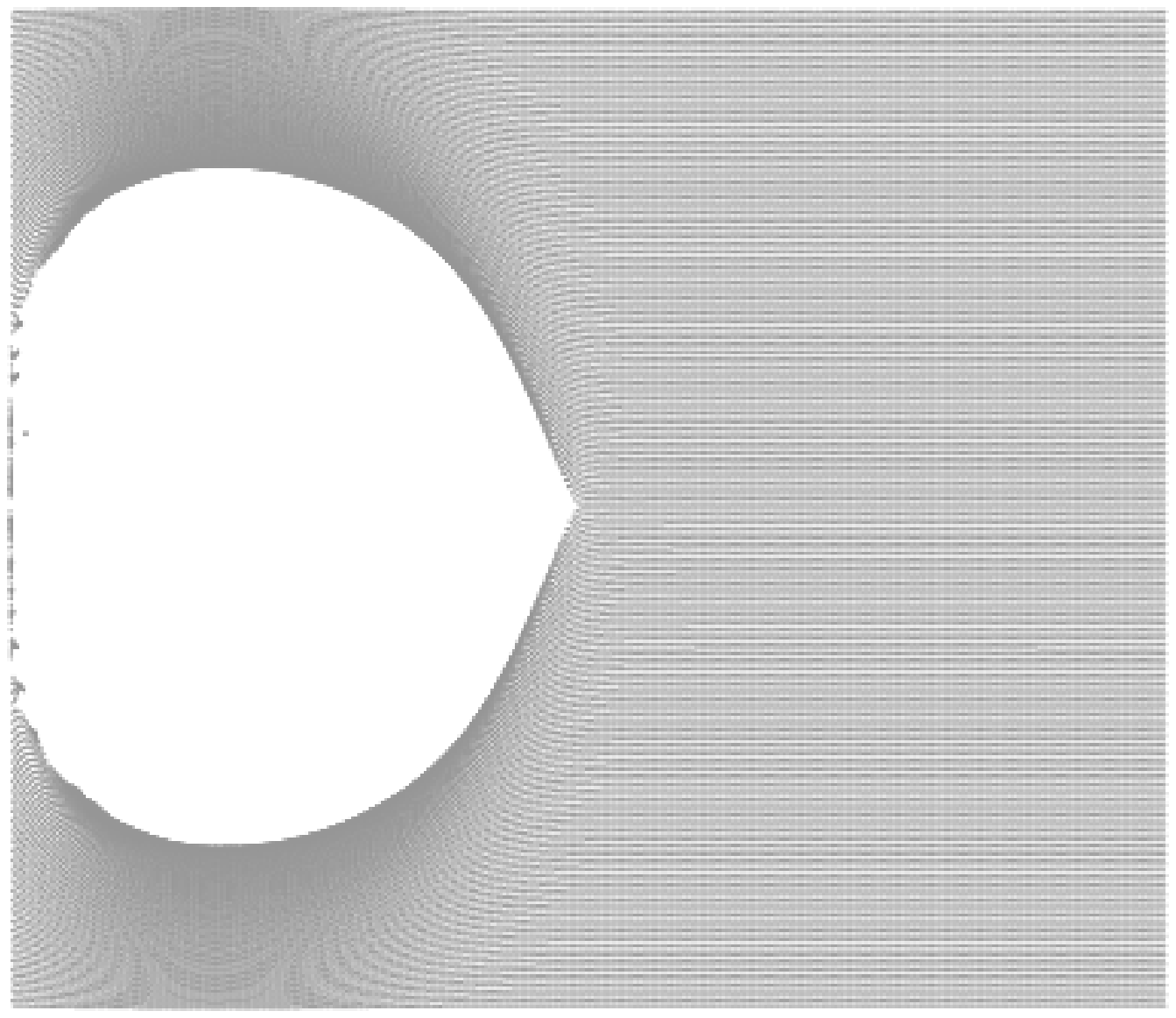}\includegraphics[%
  width=0.40\columnwidth,
  keepaspectratio]{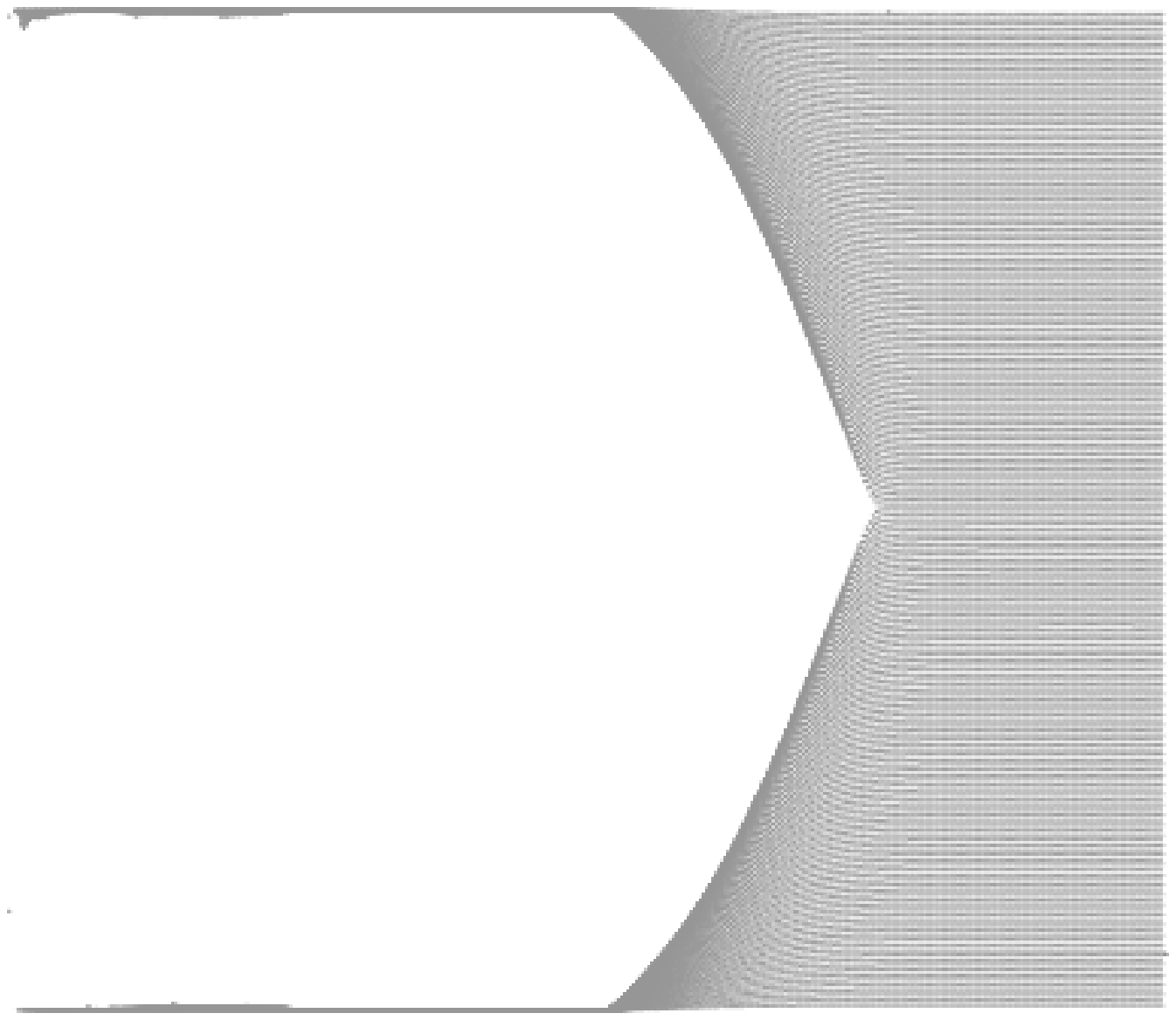}\end{center}

\caption{Four panels showing initiation of rupture in numerical sheet of rubber.
The numerical system contains around 70,000 particles, and solves
Eq. \prettyref{eq:motion}, using Eqs. \prettyref{eq:Umic} and \prettyref{eq:Umic_MR}
with $\MRa=501$m$^{2}/$s$^{2},$ $\MRb=.106,$ $\lambda_{f}=5.5,$
and $\beta=3.$ The first panel shows the initial pop, the second
shows the system 12.5 time units later, the third after 25 time units,
and the final panel after 250 time units, where the time unit is $a/\sqrt{\MRa}$
. \label{cap:Four-panels-showing}}
\end{figure}
Figure \ref{cap:Four-panels-showing} shows characteristic panels
from a numerical solution of Eq. \prettyref{eq:motion}. First, one
prepares a uniformly strained sheet, in this case with extensions
$\lambda_{x}=2.2,$ $\lambda_{y}=3.2.$ Next, two rows of particles
are selected near the left hand side of the sample: the rows are five
particles wide, and they sit right on top of one another. The upper
particles are given a large upward velocity, and the lower particles
are given a large downward velocity. This initial condition has the
effect of popping a hole in the strip. As shown in the subsequent
panels of Figure \ref{cap:Four-panels-showing}, the hole initially
develops in a circular fashion, but as it senses the upper and lower
boundaries, it begins to run sideways, and eventually turns into a
shock-like rupture front that travels in steady state indefinitely
to the right.

\begin{figure}
\begin{center}\includegraphics{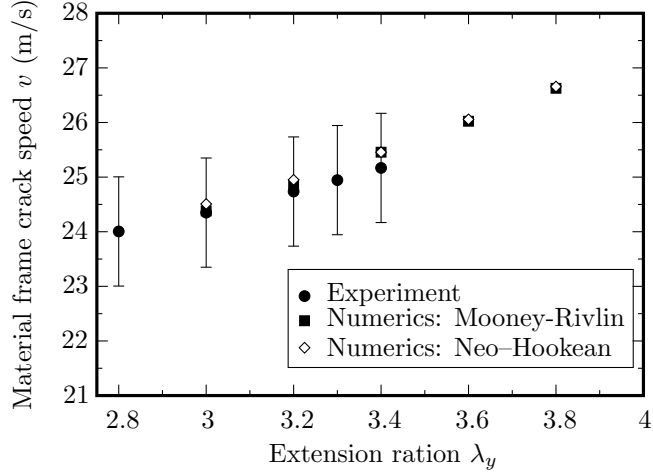}\end{center}

\caption{Comparison of experimental crack speeds and speeds in simulation.
Simulations and experiments are conducted with $\lambda_{x}=2.2,$
and a range of vertical extensions $\lambda_{y}$ .The parameters
describing the properties of the continuum nonlinear elastic theory
are $\MRa=501\ \mbox{(m/s})^{2},$ $\ \MRb=0.106$, as in Figure \ref{fig:sound_data}.
Bonds in the simulation fail when extended an amount $\lambda_{f}=5.5$
above their original length. The magnitude of Kelvin dissipation is
$\beta=3$. In addition, the figure displays results from a substantially
simplified numerical model where $\MRb=0,$ and where $E_{zz}$ is
set to zero as well. Stripping most of the complexity from the numerics
has little effect on the results.\label{cap:Comparison-of-experimental}}
\end{figure}

Figure \ref{cap:Comparison-of-experimental} shows a comparison of
experimental rupture velocities once steady state has been reached
with results from numerical simulations. The agreement is satisfactory.
However, the figure also contains the results of a different set of
simulations. In these, Eq. \prettyref{eq:motion} is solved not with
the Mooney-Rivlin energy function in Eq. \prettyref{eq:Umic_MR},
but with a much simplified Neo--Hookean energy functional\begin{equation}
\energy_{NH}(I_{1}^{i},I_{2}^{i})=c^{2}I_{1}^{i}.\label{eq:Umic_NH}\end{equation}
The velocity of ruptures described by this very simple theory is indistinguishable
from the velocity of ruptures described by the more elaborate Mooney-Rivlin
theory. This observation opens the way to accurate analytical descriptions
of the rupture of rubber, both at the continuum and discrete levels.

\section{Continuum Neo--Hookean theory\label{sec:Continuum-Neo--Hookean-theory}}

\subsection{Cracks in the Neo--Hookean theory}

The continuum Neo--Hookean energy is\begin{equation}
\energy_{NH}=\MRa I_{1}=\frac{\MRa}{2}\left[\left(\frac{\partial u^{x}}{\partial x}\right)^{2}+\left(\frac{\partial u^{x}}{\partial y}\right)^{2}+\left(\frac{\partial u^{y}}{\partial x}\right)^{2}+\left(\frac{\partial u^{y}}{\partial y}\right)^{2}\right].\end{equation}

This energy functional was first employed for the study of rubber
by \cite{Mooney1940}, and was employed for the study of fractures
by \cite{Klingbeil.66}. It has a number of interesting properties.
Because it is quadratic in displacements, it leads to a linear equation
of motion for $\vec{u,}$ which is easy to approach analytically.
Nevertheless, it describes very large displacements of rubber, and
in that sense is still a nonlinear theory. According to Eq. \prettyref{eq:MR_soundspeeds},
there is only one sound speed for this theory, \begin{equation}
c^{2}=A.\label{eq:NeoHookeanSoundSpeed}\end{equation}
The equation of motion of the Neo--Hookean theory follows from Eqs
\prettyref{eq:Fi}, \prettyref{eq:Umic}, and \prettyref{eq:motion},
and is\begin{equation}
m\ddot{u}_{i}^{\alpha}=\sum_{j\in n(i)}\frac{2mc^{2}}{3\lspace^{2}}\left(u_{ij}^{\alpha}+\beta\dot{u}_{ij}^{\alpha}\right)\theta(\lambda_{f}-u_{ij}).\label{eq:motion_NH}\end{equation}
In the continuum limit, overlooking bond rupture, one has

\begin{eqnarray}
m\ddot{u_{i}}^{\alpha} & = & \frac{2mc^{2}}{3\lspace^{2}}\sum_{j\in n(i)}(u_{j}^{\alpha}-u_{i}^{\alpha})+\beta(\dot{u}_{j}^{\alpha}-\dot{u}_{i}^{\alpha})\nonumber \\
 & = & \frac{mc^{2}}{3\lspace^{2}}\sum_{\vec{\lspace}\alpha\gamma}\frac{\partial^{2}(u^{\alpha}+\beta\dot{u}^{\alpha})}{\partial r_{\alpha}\partial r_{\gamma}}a_{ij}^{\alpha}a_{ij}^{\gamma}\label{eq:continuum_dissipation}\\
 & = & mc^{2}\nabla^{2}(u^{\alpha}+\beta\dot{u}^{\alpha})\nonumber \end{eqnarray}

For the study of fracture, there are some additional simplifications
that arise in simple geometries. Consider a semi-infinite crack moving
along the center line of an infinitely long strip as illustrated in
Figure \ref{cap:Illustration-of-reference}. The bottom of the strip
is held at height $-b\lambda_{y}$, while the top of the strip is
raised rigidly to height $\lambda_{y}b,$ where \textbf{$2b$} is
the original height of the strip. Suppose that the rubber is initially
stretched by a factor $\lambda_{x}$ everywhere in the horizontal
direction. Dropping for the moment Kelvin dissipation, the equation
of motion is

\begin{subequations}\begin{equation}
\ddot{u}^{x}=c^{2}\nabla^{2}u^{x}\end{equation}
\begin{equation}
\ddot{u}^{y}=c^{2}\nabla^{2}u^{y}\end{equation}
 \label{eq:NH_motion0}\end{subequations}

and the boundary conditions for a crack with tip at $vt$ are \begin{subequations}\begin{eqnarray}
u^{y}(x,-b) & = & -\lambda_{y}b;\  u^{y}(x,b)=\lambda_{y}b;\ \label{eq:MR_bc}\\
\partial u^{y}/\partial y\vert_{y=0} & = & 0\quad\mbox{for }x<vt\\
u^{y}(x,0) & = & 0\quad\mbox{for }x>vt\end{eqnarray}
 \end{subequations}The main point to make here is that the boundary
conditions nowhere involve $u^{x},$ nor do the equations of motion
couple $u^{x}$ and $u^{y}$. Therefore, one can take $u^{x}(x,y)=\lambda_{x}x$
for all time, and the problem reduces to one involving only $u^{y}.$
This problem is mathematically identical to the problem of crack motion
in anti-plane shear, which is discussed in textbooks. For example,
\cite[p. 127]{Broberg.99} provides the solution of a stationary crack
in this geometry, and the solution for a crack moving at steady velocity
$v$ can be obtained from the static solution by a simple change of
variables $\tilde{x}=x/\sqrt{1-v^{2}/c^{2}}.$ The solutions become
increasingly blunt as they approach the sound speed $c.$ 

I have not included Kelvin dissipation in Eq. \prettyref{eq:NH_motion0}.
The reason is that this term would destroy the conventional $\sqrt{r}$
singularity expected for cracks. If one supposes there exists a solution
where $u^{y}(r,)\sim\sqrt{r}$, then Kelvin dissipation produces an
infinite amount of energy dissipation in the vicinity of the tip\citep{Marder.IJF.04}.
We will see, however, that for supersonic solutions where the mathematical
structure near the tip is different, Kelvin dissipation is not only
permitted but required.

\subsection{Shocks in Neo--Hookean material}

I now proceed to study supersonic solutions of the Neo--Hookean theory
in the presence of dissipation. Adopting once again the geometry of
Figure \ref{cap:Illustration-of-reference}, one can conclude that
$u^{x}(x,y)=\lambda_{x}x$ at all times, and focus only upon $u^{y}.$
Since this is the only variable to consider \emph{in the following
discussion, $u$ will refer to $u^{y}.$} The vertical displacement
$u=u^{y}$ obeys the equation of motion

\begin{equation}
\ddot{u}=c^{2}\nabla^{2}u+c^{2}\beta\nabla^{2}\dot{u},\label{eq:VST1}\end{equation}
with boundary conditions at the top and bottom of the strip still
given by Eq. \prettyref{eq:MR_bc}, but now at $y=0$\begin{equation}
\frac{\partial u}{\partial y}=-\beta\frac{\partial^{2}u}{\partial t\partial y}\quad\textrm{for}\quad\  x<0;\quad u=0\quad\textrm{for}\quad x>0.\label{eq:VST2}\end{equation}
The boundary condition is obtained heuristically by discretizing the
derivatives in the $y$ direction, and eliminating the near-neighbor
interactions for $y<0.$ That is, write \begin{equation}
\nabla^{2}u\approx\begin{array}{cc}
[ & u(x+\lspace,y+u(x-\lspace,y)\\
 & +u(x,y+\lspace)+u(x,y-\lspace)\\
 & -4u(x,y)]/\lspace^{2}\hfill\end{array}.\end{equation}
On the boundary there are no particles located at $y-\lspace,$ then
one has there instead

\begin{equation}
\nabla^{2}u\approx\begin{array}{cc}
[ & u(x+\lspace,y+u(x-\lspace,y)-2u(x,y)\\
 & +\left\{ u(x,y+\lspace)-u(x,y)\right\} ]/\lspace^{2}\hfill\end{array}.\end{equation}
 The term in curly brackets must vanish, or it produces contributions
of order $1/\lspace.$ Analyzing also the last term of Eq. \prettyref{eq:VST1}
in this way produces Eq. \prettyref{eq:VST2}.

In steady state, the equation of motion and boundary condition become

\begin{equation}
v^{2}\frac{\partial^{2}u}{\partial x^{2}}=c^{2}\nabla^{2}u-v\beta c^{2}\nabla^{2}\frac{\partial u}{\partial x},\label{eq:VST3}\end{equation}
with boundary condition at $y=0$\begin{equation}
\frac{\partial u}{\partial y}=v\beta\frac{\partial^{2}u}{\partial x\partial y}\quad\textrm{for}\quad\  x<0;\quad u=0\quad\textrm{for}\quad x>0.\label{eq:VST4}\end{equation}
This system can be solved by the Wiener--Hopf technique. Consider
Eq. \prettyref{eq:VST3} for $y>0.$ Subtract out the asymptotic behavior
far ahead of the rupture, with 

\begin{equation}
w(x,y)\equiv u(x,y)-\lambda_{y}y.\label{eq:VST4.1}\end{equation}
Then\begin{equation}
v^{2}\frac{\partial^{2}w}{\partial x^{2}}=c^{2}\nabla^{2}w-v\beta c^{2}\nabla^{2}\frac{\partial w}{\partial x},\label{eq:VST4.2}\end{equation}
with boundary condition at $y=0$\begin{eqnarray}
\frac{\partial w}{\partial y}+\lambda_{y} & = & v\beta\frac{\partial^{2}w}{\partial x\partial y}\quad\textrm{for}\quad\  x<0;\label{eq:VST4.3a}\\
w & = & 0\quad\textrm{for}\quad x>0.\end{eqnarray}
Next, Fourier transform along $x$ with \begin{equation}
W(k,y)\equiv\int dx\,\exp[ikx]w(x,y).\label{eq:VST5}\end{equation}
 Then \begin{eqnarray}
-k^{2}v^{2}W & = & c^{2}(\frac{\partial^{2}}{\partial y^{2}}-k^{2})(1+ikv\beta)W\label{eq:VST6}\\
\frac{\partial^{2}W}{\partial y^{2}} & = & k^{2}\left(1-\frac{v^{2}}{c^{2}(1+ikv\beta)}\right)W\label{eq:VST7}\\
\Rightarrow W & = & W_{0}(k)e^{-gy},\label{eq:VST8}\\
g(k) & = & k\sqrt{\frac{c^{2}-v^{2}+ikv\beta c^{2}}{c^{2}(1+ikv\beta)}}\textrm{with}\ \Re(g)>0\end{eqnarray}
When $v<c,$ in order to insure that the real part of $g$ is positive,
one should write it as\[
g(k)=\sqrt{k^{2}+\epsilon^{2}}\sqrt{\frac{c^{2}-v^{2}+ikv\beta c^{2}}{c^{2}(1+ikv\beta)}},\]
where $\epsilon$ is small. However, when $v>c,$ one must write instead\begin{equation}
g(k)=ik\sqrt{\frac{v^{2}-c^{2}-ikv\beta c^{2}}{c^{2}(1+ikv\beta)}}.\label{eq:VST10}\end{equation}
Note that there is a branch of $g(k)$ with positive real part everywhere
as $k$ moves along the real axis. The problem reduces to finding
$W_{0}(k).$ This function may be determined from the boundary conditions.
To do so, write\[
w_{0}(x)=w(x,y=0)\]
 \begin{equation}
W_{0}(k)=\int dx\  w_{0}(x)e^{ikx}=\int_{-\infty}^{0}dx\: w_{0}(x)e^{ikx}\equiv W_{0}^{-}(k).\label{eq:W0eq}\end{equation}
The superscript $-$ indicates that $W_{0}^{-}$ has no poles in the
lower half plane. Next, introducing a convergence factor $\exp[-\epsilon|x|]$
to keep the constant $\lambda_{y}$ under control, sending $\epsilon$
to zero at the end of the calculation, write the boundary condition
\prettyref{eq:VST4.3a} as\begin{eqnarray}
 &  & \int dx\ \left(\frac{\partial w}{\partial y}+\lambda_{y}e^{-\epsilon|x|}-v\beta\frac{\partial^{2}w}{\partial x\partial y}\right)e^{ikx}=-gW_{0}+\frac{\lambda_{y}}{\epsilon-ik}+\frac{\lambda_{y}}{\epsilon+ik}-v\beta ikgW_{0}\nonumber \\
 & = & \int_{0}^{\infty}dx\ \left(\frac{\partial w}{\partial y}+\lambda_{y}e^{-\epsilon|x|}-v\beta\frac{\partial^{2}w}{\partial x\partial y}\right)e^{ikx}\equiv Q^{+}(k),\label{eq:bc_result}\end{eqnarray}
where the superscript $+$ indicates that $Q^{+}$ has no poles in
the upper half plane. Therefore, using Eq. \prettyref{eq:W0eq} one
can write\begin{equation}
-g(1+iv\beta k)W_{0}^{-}+\frac{\lambda_{y}}{\epsilon-ik}+\frac{\lambda_{y}}{\epsilon+ik}=Q^{+}.\label{eq:VST15}\end{equation}
Define\begin{eqnarray}
G(k) & = & g(k)(1+iv\beta k)/ik\nonumber \\
 & = & \sqrt{(1+iv\beta k)(v^{2}/c^{2}-1-ivk\beta)}=G^{+}(k)G^{-}(k),\end{eqnarray}
\begin{eqnarray}
\mbox{with}\quad G^{+}(k) & = & \sqrt{v^{2}/c^{2}-1-ivk\beta};\quad G^{-}(k)=\sqrt{1+iv\beta k}.\end{eqnarray}
Note that $G^{+}$ is free of poles or zeroes in the upper half plane
(it has a zero in the lower half plane) while $G^{-}$ is free of
poles or zeroes in the lower half plane (it has a zero in the upper
half plane). On the real axis, one takes the branch of both $G^{+}$
and $G^{-}$ that is positive when $k=0;$ this ensures that the real
part of $g(k)$ is positive as required. Therefore, write Eq. \prettyref{eq:VST15}
as \begin{eqnarray}
Q^{+} & = & -ikG^{+}G^{-}W_{0}^{-}+\frac{\lambda_{y}}{\epsilon-ik}+\frac{\lambda_{y}}{\epsilon+ik}\nonumber \\
\Rightarrow &  & \frac{Q^{+}}{G^{+}}-\frac{\lambda_{y}}{\epsilon-ik}\frac{1}{G^{+}(0)}=\frac{\lambda_{y}}{(\epsilon+ik)G^{+}(0)}-ikG^{-}W_{0}^{-}.\label{eq:VST20}\end{eqnarray}
The two sides of Eq. \ref{eq:VST20} have poles on opposite sides
of the real axis, and must separately equal a constant. If the constant
is nonzero, then $w_{0}$ will be discontinuous at the origin, while
it if is zero, $w_{0}$ is continuous although $\partial w_{0}/\partial x$
is discontinuous. Therefore, take the constant to be zero. One has\begin{eqnarray}
ikW_{0}^{-} & = & \frac{\lambda_{y}}{(\epsilon+ik)\sqrt{1+iv\beta k}\sqrt{v^{2}/c^{2}-1}}\\
 & \Rightarrow & \sqrt{v^{2}/c^{2}-1}\frac{\partial^{2}w_{0}}{\partial x^{2}}=\int\frac{dk}{2\pi}\,\frac{\lambda_{y}e^{-ikx}}{\sqrt{1+iv\beta k}}=\int_{0}^{\infty}\frac{dk}{2\pi}\:\frac{\lambda_{y}e^{-ikx}}{\sqrt{1+iv\beta k}}+{\rm c.c.}\nonumber \end{eqnarray}
There is a nonzero result only when $x<0.$ In this case, one must
deform the contour so that $k$ travels along the positive imaginary
axis; $k\rightarrow ik'.$ The branch of the square root is one that
has positive imaginary part on the right side of the imaginary axis
as one deforms the contour. Making this change of variables, one has\begin{equation}
\int_{0}^{\infty}\frac{i\, dk'}{2\pi}\:\frac{\lambda_{y}e^{k'x}}{\sqrt{1-v\beta k'}}+{\rm c.c.}.\end{equation}
From 0 to $1/\beta v$, the integrand is purely imaginary and cancels
with the complex conjugate. For the remainder of the contour, the
square root is taken on the branch with positive imaginary part and
gives\begin{eqnarray}
 &  & \int_{1/v\beta}^{\infty}\frac{dk'}{2\pi}\:\frac{\lambda_{y}e^{k'x}}{\sqrt{v\beta k'-1}}+{\rm c.c.}=2\int_{0}^{\infty}\frac{dk'}{2\pi}\:\frac{\lambda_{y}e^{k'x+x/v\beta}}{\sqrt{v\beta k'}}\nonumber \\
 & = & \frac{\lambda_{y}e^{x/v\beta}}{\sqrt{-\pi v\beta x}}.\end{eqnarray}
Thus one has\begin{eqnarray}
\sqrt{v^{2}/c^{2}-1}\frac{\partial^{2}w_{0}}{\partial x^{2}} & = & \begin{cases}
\frac{\lambda_{y}e^{x/v\beta}}{\sqrt{-\pi v\beta x}} & \textrm{for }x<0\\
0 & \textrm{else}\end{cases}\\
\Rightarrow\sqrt{v^{2}/c^{2}-1}\frac{\partial w_{0}}{\partial x} & = & -\int_{x}^{0}dx'\:\frac{\lambda_{y}e^{x/v\beta}}{\sqrt{-\pi v\beta x}}.\end{eqnarray}
In particular, the slope of the face of the rupture as $x\rightarrow-\infty$
is given by\begin{equation}
\frac{\partial w_{0}(-\infty)}{\partial x}=-\frac{\lambda_{y}}{\sqrt{v^{2}/c^{2}-1}}.\end{equation}
In the lab, the slope of the crack face will be\begin{equation}
-\frac{\lambda_{y}}{\lambda_{x}\sqrt{v^{2}/c^{2}-1}}\label{eq:lab_face_slope}\end{equation}
and the opening angle $\theta$ will be\begin{equation}
\theta=2\,\tan^{-1}\left[\frac{\lambda_{y}}{\lambda_{x}\sqrt{v^{2}/c^{2}-1}}\right].\end{equation}
A peculiar aspect of the neo-Hookean theory once terms involving $E_{zz}$
have been discarded is that it describes material which in its lowest
energy state shrinks down into a point. This unphysical feature provides
an advantage in this calculation, since it means that the slope of
the crack face is also the same as the slope of the line along which
material begins to deform as the rupture approaches. As predicted
in Section \ref{sec:Elementary-Theory}, Eq. \prettyref{eq:lab_face_slope}
is exactly 

In order to determine the velocity of the rupture, one needs a criterion
to describe when material fails. Consider a bond lying on the center
line just before the tip of the rupture which in the material frame
points along $(1/2\ \ \sqrt{3}/2).$ In the numerical model studied
in this paper, the bonds that snap are all of this sort. Experiments
are carried out in amorphous materials, and it would remain to be
shown that this type of bond is sufficiently representative of those
that snap. Within a continuum framework, it is natural to suppose
that this bond snaps when\begin{equation}
\lambda_{f}^{2}=\frac{1}{4}\lambda_{x}^{2}+\frac{3}{4}\left(\left.\frac{\partial u}{\partial y}\right|_{{x=0\atop y=0}}\right)^{2}.\label{eq:VST31}\end{equation}
It is easy to come up with more elaborate criteria, but this one has
the virtue of simplicity, and accounts reasonably well both for experimental
and numerical results. In order to compute the quantity on the right
side of Eq. \prettyref{eq:VST31}, note that\begin{eqnarray}
\int dx\left.\frac{\partial w}{\partial y}\right|_{y=0}e^{ikx} & = & -gW_{0}=-ik\frac{G^{+}}{G^{-}}W_{0}\nonumber \\
 & = & -\frac{\lambda_{y}G^{+}}{(1+ikv\beta)(\epsilon+ik)\sqrt{{v^{2}/c^{2}-1}}}\end{eqnarray}
\begin{equation}
\left.\frac{\partial w}{\partial y}\right|_{y=0}=-\int\frac{dk}{2\pi}\: e^{-ikx}\frac{\lambda_{y}\sqrt{v^{2}/c^{2}-1-ivk\beta}}{(1+ikv\beta)(\epsilon+ik)\sqrt{{v^{2}/c^{2}-1}}}.\end{equation}
For $x<0,$ one must close the contour in the upper half plane, where
there are two poles, one at $i/v\beta,$ and one at $i\epsilon$.
These contribute\begin{eqnarray}
\Rightarrow\left.\frac{\partial w}{\partial y}\right|_{y=0} & = & -\frac{2\pi i}{2\pi}\frac{\lambda_{y}e^{x/v\beta}G^{+}(i/v\beta)}{iv\beta(-1/v\beta)\sqrt{{v^{2}/c^{2}-1}}}-\frac{2\pi i}{2\pi}\frac{\lambda_{y}G^{+}(0)}{i\sqrt{v^{2}/c^{2}-1}}\nonumber \\
 & = & \frac{\lambda_{y}e^{x/v\beta}v/c}{\sqrt{v^{2}/c^{2}-1}}-\lambda_{y}\Rightarrow\left.\frac{\partial u}{\partial y}\right|_{y=0}=\frac{\lambda_{y}e^{x/v\beta}v/c}{\sqrt{v^{2}/c^{2}-1}},\end{eqnarray}

and in particular at $x=0$ one has\begin{equation}
\left.\frac{\partial u}{\partial y}\right|_{{{x=0\atop y=0}}}=\frac{\lambda_{y}v/c}{\sqrt{v^{2}/c^{2}-1}}.\label{eq:dudy0}\end{equation}

Similarly, for $x>0$ one finds that\begin{eqnarray}
\frac{\partial u}{\partial y} & \Big\vert_{y=0}= & \lambda_{y}+\int\frac{\lambda_{y}dt}{\pi}\frac{t^{2}e^{-(t^{2}+v^{2}/c^{2}-1)x/v\beta}}{(t^{2}+\frac{v^{2}}{c^{2}})(t^{2}+\frac{v^{2}}{c^{2}}-1)\sqrt{\frac{v^{2}}{c^{2}}-1}}.\end{eqnarray}

It is also interesting to compute the vertical stress ahead of the
rupture, which is \begin{eqnarray}
\sigma_{y}/\rho c^{2}\Big\vert_{y=0} & = & \partial u/\partial y+\beta\partial\dot{u}/\partial y\nonumber \\
 & = & \lambda_{y}+\frac{\lambda_{y}e^{-(v^{2}/c^{2}-1)x/(\beta v)}}{\sqrt{\pi x/(\beta v)}\sqrt{v^{2}/c^{2}-1}}\nonumber \\
 &  & -\lambda_{y}\sqrt{v^{2}/c^{2}-1}\int_{x}^{\infty}\frac{dx'}{v\beta}\,\frac{e^{-(v^{2}/c^{2}-1)x'/(\beta v)}}{\sqrt{\pi x/(\beta v)}}.\end{eqnarray}
 Note that while the displacement gradient and strain are finite in
front of the rupture, the stress does have an inverse square root
singularity at the origin. This singularity is due to the Kelvin dissipation,
and does not indicate that there is a finite energy flux to the tip
as in conventional fracture. 

Returning now to Eq. \prettyref{eq:dudy0}, the rupture criterion
\prettyref{eq:VST31} becomes\begin{eqnarray}
\lambda_{f}^{2} & = & \frac{1}{4}\lambda_{x}^{2}+\frac{3}{4}\frac{\lambda_{y}^{2}v^{2}/c^{2}}{v^{2}/c^{2}-1}\\
\Rightarrow\tilde{\lambda}_{y} & \equiv & \frac{\lambda_{y}}{\sqrt{(4\lambda_{f}^{2}-\lambda_{x}^{2})/3}}=\sqrt{1-c^{2}/v^{2}}\label{eq:simple_result}\end{eqnarray}
Note that this expression predicts a specific way of assembling samples
with different values of $\lambda_{x}$ and $\lambda_{y}$ that all
should travel at the same speed $v.$ For the exact solution of a
discrete theory presented in the next section, the final result is
of exactly this same form, with the same quantity $\tilde{\lambda}_{y}$
appearing, but related to a more complicated function of $v/c$.

\subsection{Disintegration of back face\label{sub:Disintegration-of-back}}

\begin{figure}
\begin{center}\includegraphics[%
  bb=150bp -1bp 300bp 197bp,
  clip,
  width=0.40\columnwidth,
  keepaspectratio]{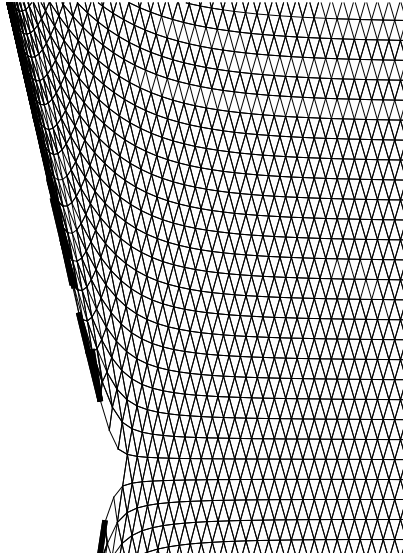}\end{center}

\caption{Close-up view of top half of rupture with $\lambda_{y}=3.4,$ $\lambda_{x}=2.2,$
$\lambda_{f}=5.5$, $\beta=3,$ traveling at $v/c=1.09.$ Bonds stretched
beyond $\lambda_{f}$ are drawn much thicker than other bonds. As
predicted by \prettyref{eq:lambda_back}, a line of bonds along the
back edge of the rupture has been stretched beyond $\lambda_{f},$
and they remain intact only because of the rule that toughens all
bonds connected to a node where at least one bond has shrunk below
1.5 times its equilibrium length. Note that several bonds off the
main crack line snap as the tip progresses, but not enough to destroy
the integrity of the material. Without some sort of toughening rule,
however, the entire back edge of the rupture must disintegrate.\label{cap:Close-up-view-of}}
\end{figure}

To obtain a final lesson from the continuum solutions, return to Eq.
\prettyref{eq:lab_face_slope}. Consider two mass points that before
the arrival of the rupture lie on the central axis at horizontal distance
$dx$ from each other. According to this expression, on the back face
of the rupture, they are now separated by the squared distance\[
dx^{2}+dx^{2}\frac{\lambda_{y}^{2}}{\lambda_{x}^{2}(v^{2}/c^{2}-1)},\]
which means that material along the back end of the rupture is stretched
by amount $\lambda_{{\rm back}}$ where\[
\lambda_{{\rm back}}^{2}=\lambda_{x}^{2}+\frac{\lambda_{y}^{2}}{(v^{2}/c^{2}-1)}.\]
Employing Eq. \prettyref{eq:simple_result}, one has that 

\begin{equation}
\lambda_{{\rm back}}^{2}=\lambda_{x}^{2}(1-\frac{c^{2}}{3v^{2}})+\frac{4}{3}\frac{c^{2}}{v^{2}}\lambda_{f}^{2}.\label{eq:lambda_back}\end{equation}
Inspection of Eq. \prettyref{eq:lambda_back} makes it plausible that
extensions along the back end of the rupture can be greater than $\lambda_{f}:$
that is, they are generally greater than the extension at which rubber
near the tip is supposed to give way. This is the reason that material
must toughen behind the rupture tip. Otherwise, no steady solution
is possible and the back end of the rupture disintegrates. To emphasize
this point, Figure \ref{cap:Close-up-view-of} shows a numerical rupture
solution with bonds in bold when they have stretched beyond $\lambda_{f}.$
As predicted by Eq. \prettyref{eq:lambda_back}, the entire back surface
of the rupture is in this state.

This calculation explains the need for the toughening rule described
after Eq. \prettyref{eq:motion}. Without such a rule, no supersonic
solutions appear in numerical calculations. With it they become generic.

\section{Discrete Neo-Hookean theory\label{sec:Discrete-Neo-Hookean-theory}}

\begin{figure}[th]
\begin{center}\includegraphics[%
  width=0.40\columnwidth,
  keepaspectratio]{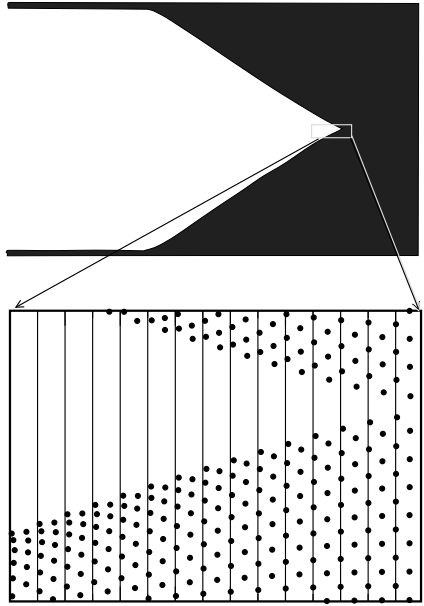}\end{center}

\caption{Snapshot of supersonic rupture in Neo--Hookean theory taken from
numerical time evolution of Eq. \prettyref{eq:motion_NH}. Note that
particles do move almost purely vertically, as shown by comparing
particle positions with the vertical lines.\label{cap:Supersonic-crack-in}}
\end{figure}

The main weakness in the continuum theory of the previous Section
is that the rupture criterion is approximate. It is possible to do
much better, since one can solve analytically the equations of motion
for the discrete Neo--Hookean theory given by Eq. \prettyref{eq:motion_NH}.
Recall that this equation was obtained with the following assumptions:

\begin{enumerate}
\item The coefficient $B$ in \prettyref{eq:Umic_MR} vanishes.
\item $E_{zz}$ can be set to zero. Because of this assumption, the energy
functional is quadratic, and the equations of motion are linear.
\item Mass points move only vertically. In fact, the horizontal forces on
all mass points balance, except during a brief time when only one
of the bonds has snapped for a mass point lying on the crack line.
Comparison of analytical solutions with direct numerical integration
of the equations of motion indicates that errors introduced by this
approximation are on the order of no more than one percent, and a
snapshot from a numerical solution of the Neo--Hookean theory shown
in Figure \ref{cap:Supersonic-crack-in} demonstrates that this approximation
is obeyed well in the vicinity of the tip.
\end{enumerate}

The calculation of steady states for Eq. \prettyref{eq:motion_NH}
is lengthy, and relegated to Appendix \ref{app:discrete}. The final
results are as follows:

Begin by specifying the dimensionless velocity and damping\begin{equation}
\tilde{v}=v/c,\,\,\tilde{\beta}=\beta c/\lspace,\end{equation}
 and compute

\begin{equation}
\zeta=\frac{3-\cos(\omega/\tilde{v})-3\omega^{2}/[4(1-i\tilde{\beta}\omega)]}{2\cos(\omega/2\tilde{v})}\label{TL8}\end{equation}

\begin{equation}
\phi=\zeta+\sqrt{\zeta^{2}-1}\mbox{with }\quad\textrm{abs}(\phi)>1,\label{TL9}\end{equation}
\[
F(\omega)=\left\{ \frac{\phi^{[N-1]}-\phi^{-[N-1]}}{\phi^{N}-\phi^{-N}}-2\zeta\right\} \cos(\omega/2\tilde{v})+1,\]

\begin{eqnarray}
\mbox{and}\quad Q(\omega) & = & \frac{F}{F-1-\cos(\omega/2\tilde{v})}.\end{eqnarray}
Then the scaled extension $\tilde{\lambda}_{y}$ defined in Eq. \prettyref{eq:simple_result}
by $\tilde{\lambda_{y}}=\lambda_{y}/\sqrt{(4\lambda_{f}^{2}-\lambda_{x}^{2})/3}$
is given as a function of $\tilde{v}$ by \begin{eqnarray}
\tilde{\lambda}_{y} & = & \frac{1}{\sqrt{2N+1}}\exp\left[-\int\frac{d\omega'}{4\pi}\left\{ \frac{\left[\ln Q(\omega')-\overline{\ln Q(\omega')}\right]}{i\omega'(1+\tilde{\beta}^{2}\omega'^{2})}+\frac{\tilde{\beta}\ln|Q(\omega')|^{2}}{1+\tilde{\beta}^{2}\omega'^{2}}\right\} \right].\label{eq:result}\end{eqnarray}

\section{Results \label{sec:Comparison-with-experiment}}

\subsection{Macroscopic Limit}

\begin{figure}
\begin{center}\includegraphics{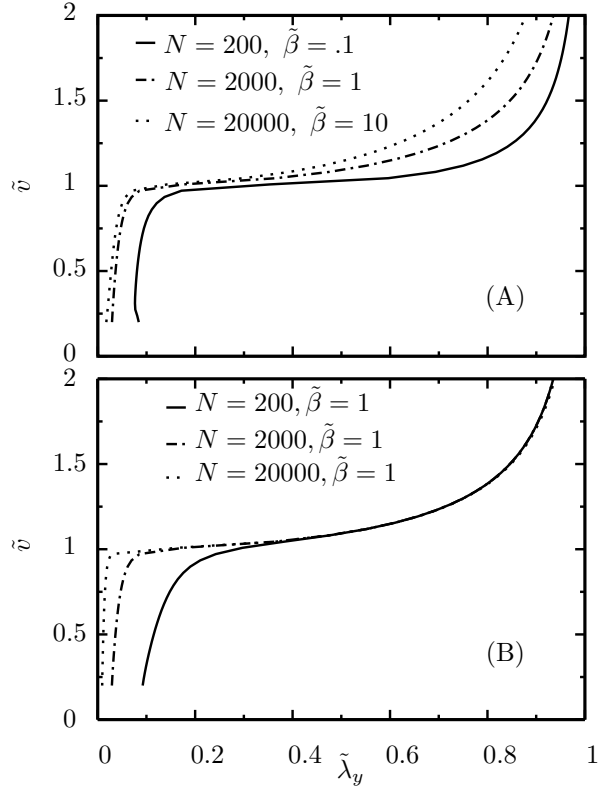}\end{center}

\caption{Sequence of velocity versus loading curves showing approaches to
the continuum limit. Curves in (A) correspond to systems of fixed
height and fixed continuum dissipation $\beta$, but sending the lattice
spacing $\lspace=L/N$ to zero. This is achieved through solutions
of Eq. \prettyref{eq:result} where $N$ increases and $\tilde{\beta}=\beta c/\lspace$
scales as $N.$ As $\lspace$ goes to zero, the subsonic branch of
solutions approaches a definite limiting value, but the velocity of
supersonic solutions increases continually as $N$ increases. Note
that on this approach to the continuum limit, the fracture energy
diminishes as $\lspace^{2}.$ Curves in (B) correspond to systems
of fixed lattice spacing and fixed continuum dissipation $\beta$.
As the height $L$ goes to infinity, the supersonic solutions approach
a limiting value, but the branch of subsonic solutions is squeezed
into a smaller and smaller region near the origin. \label{cap:Effect-of-varying}}
\end{figure}

\begin{figure}
\begin{center}\includegraphics{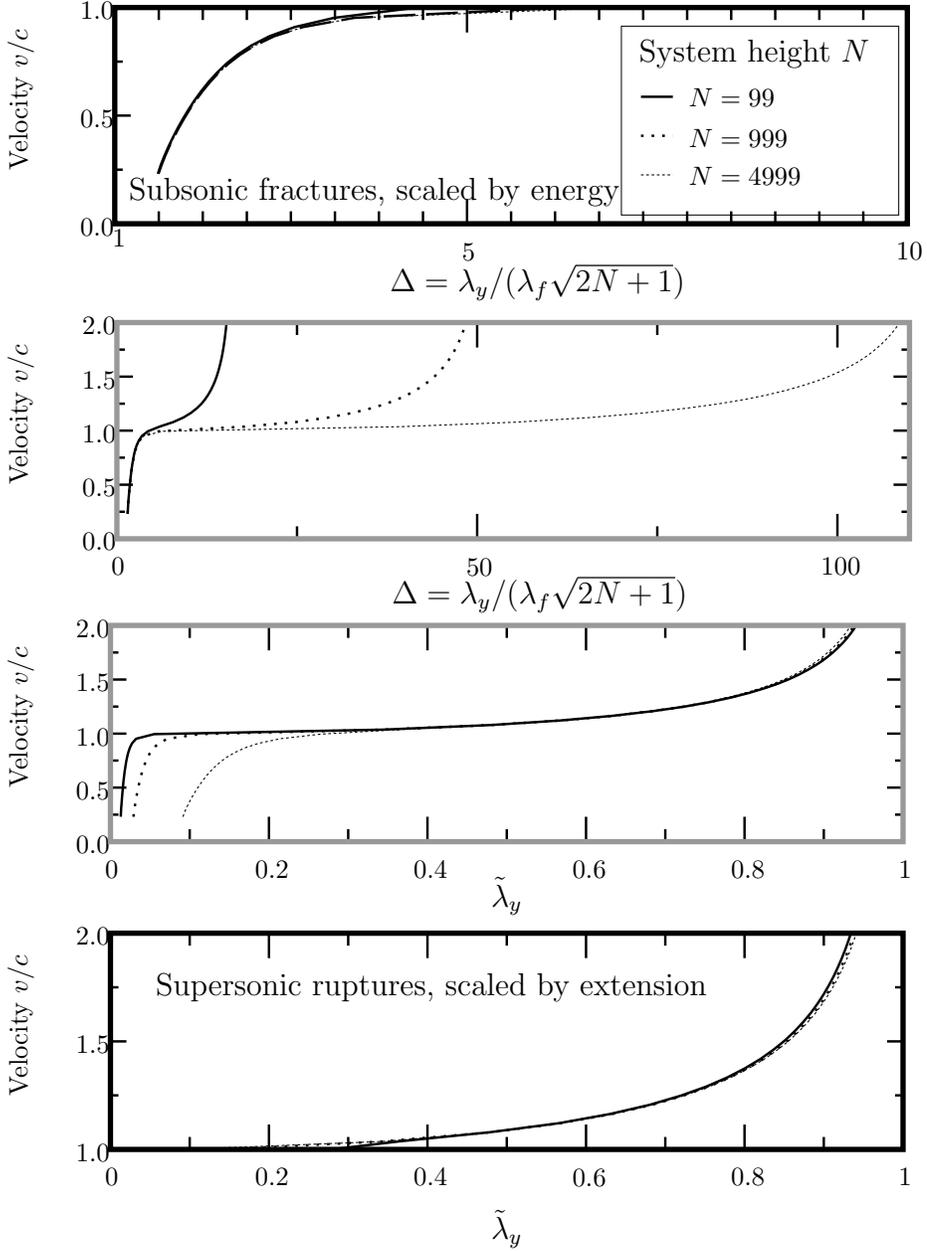}\end{center}

\caption{Four different views of Neo--Hookean crack velocities, showing that
depending upon how they are scaled and displayed, one focuses either
upon conventional subsonic fractures, or supersonic ruptures. The
definition of $\tilde{\lambda_{y}}$ is given in Eq. \prettyref{eq:simple_result}.
In the limit of infinite system height $N$, the two different types
of solutions are separated by an infinitely long plateau at the wave
speed. \label{cap:Four-different-views}}
\end{figure}

Eq. \prettyref{eq:result} provides a complete expression for the
collection of extensions $\lambda_{x}$ and $\lambda_{y}$ that result
in a rupture moving at velocity $v.$ Apart from the scaled velocity
$\tilde{v}=v/c,$ the result depends upon three parameters; the system
height $N$, the extension $\lambda_{f}$ at which bonds snap, and
the coefficient of Kelvin dissipation $\tilde{\beta}=c\beta/\lspace.$
One can now search for the conditions under which one should expect
subsonic fractures, and the conditions under which one should expect
supersonic ruptures. First, consider systems of fixed height and fixed
level of dissipation $\beta$ as the lattice spacing tends to zero.
This situation is described by fixing $L$ and taking the limit as
$N\rightarrow\infty$ of Eq. \prettyref{eq:result} with $\tilde{\beta}=c\beta N/L$
also scaling as $N.$ Figure \ref{cap:Effect-of-varying} (A) shows
that in this limit, there is a narrow band of subsonic solutions followed
by a broad band of supersonic solutions. Note that since the failure
extension $\lambda_{f}$ remains fixed while the lattice spacing $\lspace=L/N$
vanishes, the fracture energy vanishes as $\lspace^{2}$ during this
limiting procedure. Thus, this limiting procedure, which at first
seems the most sensible, corresponds to something physically rather
odd. Alternatively, one can set $\beta$ to a constant and send $N$
to $\infty$ so that the sample becomes infinitely high. In this limit,
plotting solutions versus $\tilde{\lambda_{y}}$, the subsonic ruptures
disappear, and only supersonic solutions survive, as shown in Fig.
\ref{cap:Effect-of-varying} (B). However, ones conclusions about
the true nature of this macroscopic limit depend upon how one scales
the solutions, as illustrated in Fig. \ref{cap:Four-different-views}.
For a system of any given height, there are both subsonic and supersonic
solutions. The subsonic solutions are found at small strain, and as
$N$ becomes large, the range of extensions $\tilde{\lambda}_{y}$
that produces them becomes progressively smaller. There is a plateau
near the wave speed that becomes wider and wider as $N$ increases.
Finally, supersonic ruptures appear for extensions $\lambda_{y}$
on the order of $\lambda_{f}.$ The point to emphasize is that depending
how extensions are scaled, either the supersonic or subsonic branches
can be viewed as the macroscopic limit.  In most brittle materials
it is impossible for cracks to reach the wave speed because they become
unstable to side-branching before this point is reached \citep{Fineberg.99}.
One of the things that appears to make rubber different is that the
ruptures are so stable that it is possible for them to pass the wave
speed and move beyond it without instabilities intervening. 

\subsection{Comparison with Experiment}

To close this investigation, I compare the results with experiments
on rupture of rubber sheets. It was already demonstrated in Figure
\ref{fig:sound_data} that the Mooney-Rivlin theory adequately captures
the variation of sound speed with extension. The remaining two quantities
measured by \citet{Petersan.04} are rupture speeds and opening angles.
Before making the comparison, two reasons to view the comparison with
a bit of skepticism should be noted. First, rubber is an entangled
polymer network, not a triangular lattice. Second, although the Kelvin
dissipation proportional to $\beta$ plays a very important role in
the theory, no estimate of its value from experiment has been provided.
The reason is that dissipation in real amorphous solids is not of
the form employed here, nor is there any simple way to correct the
deficiency. The spatial decay rate of sound waves in rubber over frequencies
ranging from kilohertz to megahertz is almost perfectly linear in
$\omega$ \citep{Mott2002}:\begin{equation}
\alpha=C\omega,\label{eq:aw}\end{equation}
where $C$ is a constant. However, given the form of Kelvin dissipation
employed in this paper, sound decays at the rate \begin{equation}
\alpha=\mbox{Im}(k)=\mbox{Im}\left[\omega/\left(c\sqrt{1-i\beta\omega}\right)\right].\label{eq:my_dissipation}\end{equation}
It is impossible to find a value of $\beta$ that makes Eq. \prettyref{eq:my_dissipation}
a good fit to Eq. \prettyref{eq:aw}. A more realistic rule for Kelvin
dissipation would provide a frequency-dependent sound speed according
to\begin{equation}
\frac{1}{c^{2}(\omega)}=\frac{1}{c_{\infty}^{2}}+\left(\frac{1}{c_{0}^{2}}-\frac{1}{c_{\infty}^{2}}\right)\frac{1}{1-i\beta\omega};\end{equation}
the form of dissipation used in Eq. \prettyref{eq:VST1} corresponds
to sending to the high-frequency sound speed $c_{\infty}$ to infinity.
However, the frequency dependence of sound attenuation does not resemble
experiment any better after inclusion of $c_{\infty}.$ Thus I will
simply continue to use the simplest form of Kelvin dissipation, as
it is familiar and conventional \citep{Fradkin2003} and take the
dimensionless measure of dissipation, $\tilde{\beta}=\beta c/\lspace$
to be of order unity. Fortunately, none of the final results depend
much on the value of $\tilde{\beta}$. 

Figure \ref{cap:Comparison-of-theory,} assembles experimental and
theoretical results. According to the theory for triangular lattices,
samples with extensions $\lambda_{x}$ and $\lambda_{y}$ depend only
upon the scaled variable $\tilde{\lambda}_{y}$ given by Eq. \prettyref{eq:simple_result}.
This scaling of the velocity is compatible with all the data. Thirteen
experimental trials where ruptures ran straight collapse onto five
points, with rather little variation in the scaled velocity. The scatter
in the data is rather large, and therefore consistent both with the
simplified results of Section \ref{sec:Continuum-Neo--Hookean-theory},
as well as the more elaborate results of Section \ref{sec:Discrete-Neo-Hookean-theory}.
The figure also shows a comparison of direct integration of the equations
of motion, Eq. \prettyref{eq:motion}. For the equations of motion
in this figure, the Mooney--Rivlin parameter $\MRb$ has been set
to zero, but $E_{zz}$ has not been eliminated from Eq. \prettyref{eq:w}.
Agreement with the analytical results from Eq. \prettyref{eq:result}
is excellent, showing that details of how rubber relaxes behind the
tip of the rupture do not have much effect on the dynamics. As already
shown in Figure \ref{cap:Comparison-of-experimental}, rupture speeds
are not measurably affected by including $\MRb.$ Therefore, the analytical
results of Eq. \prettyref{eq:result} capture rupture speeds, experimental
and numerical, rather completely. The simple result of Eq. \prettyref{eq:simple_result}
is adequate for a first pass. In fact, the value $\lambda_{f}=5.5$
of the bond failure extension was obtained by fitting \prettyref{eq:simple_result}
to the experimental data, and this value of $\lambda_{f}$ was then
used unchanged in all numerical runs. 

\begin{figure}
\begin{center}\includegraphics{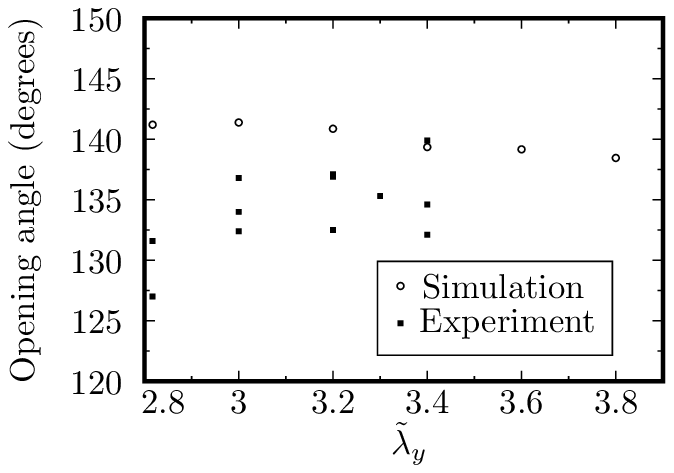}\end{center}

\caption{Rupture opening angles obtained from simulations and compared with
experiment. The agreement is not very satisfactory; the experimental
results are widely scattered, but lie systematically below the numerical
predictions. For prediction of this quantity, it may be that treating
rubber as a triangular lattice is not adequate.\label{cap:Rupure-opening-angles}}
\end{figure}

Finally, Figure \ref{cap:Rupure-opening-angles} compares experimental
results for crack opening angles with predictions based upon numerical
solutions of the most realistic numerical system, \prettyref{eq:motion}.
That is, the nonlinear terms from $E_{zz}$ that appear as rubber
shrinks towards its equilibrium relaxed state, and Rivlin's nonlinear
contribution to the Mooney-Rivlin energy are all included. Agreement
between theory and experiment for the opening angle is still not completely
satisfactory. The experimental points are widely scattered, indicating
that the reduction to the variable $\tilde{\lambda_{y}}$ may not
be appropriate, and experimental values lie systematically below theoretical
predictions. Either the simplistic form of the dissipation, or the
simplistic triangular microstructure might be to blame for this discrepancy.

There is one final potential discrepancy with experiment that should
be mentioned. According to the theory, the dynamical solutions do
include subsonic ruptures at small extensions. Many rubbers are well
known to creep \citep{Hui2003}, and tear slowly in trouser tests,
but in our biaxially loaded samples of natural latex rubber we never
observed cracks to creep, or to travel slower than the sound speed
at all.

\begin{figure}[t]
\begin{center}\includegraphics{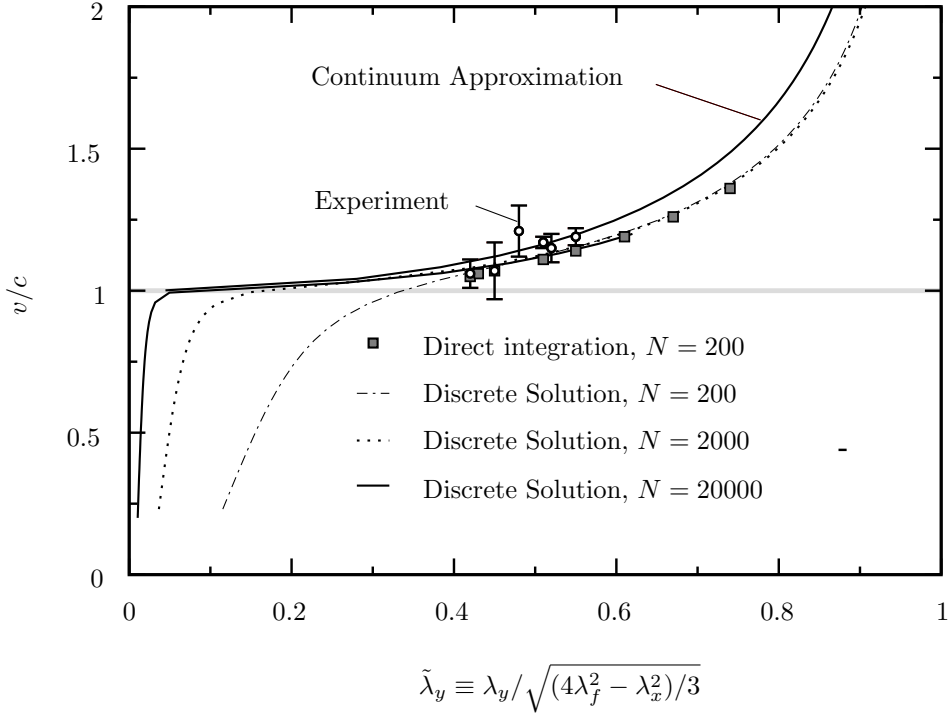}\end{center}

\caption{Comparison of theory, experiment, and numerics for rubber rupture
velocities. Experimental velocities are scaled by $c\lambda_{x}$,
with $c=$22 m/s, while the vertical extension $\lambda_{y}$ is scaled
by $\sqrt{(4\lambda_{f}^{2}-\lambda_{x}^{2})/3}$. The continuum approximation
is given in Eq. \prettyref{eq:simple_result}. Direct integration
of Eq. \prettyref{eq:motion} is carried out in triangular lattices
$N=$200 rows high in the Neo--Hookean limit where $\MRb=0,$ with
Kelvin dissipation $\beta=3,$ and retaining $E_{zz}$ as in Eq. \prettyref{eq:w}.
The Discrete  Solution is an exact solution of the same system using
the Wiener-Hopf technique,with the three differences. First, $E_{zz}$
is neglected in the analytical solution. Second, the analytical system
is infinitely long in the horizontal direction, while the numerical
system is finite. Third, in the numerical system there is a brief
time when only one of two crack-line bonds has snapped, and horizontal
forces on crack-line atoms do not balance to zero, while in the analytical
solutions, all forces in the horizontal direction are ignored. Analytical
solutions for systems both 200 and 2000 rows high are displayed to
show how the continuum limit is achieved. \label{cap:Comparison-of-theory,}}
\end{figure}

\section{Conclusions}

The main points established by this theory for the rupture of rubber
are the following:

\begin{enumerate}
\item The rupture of rubber is a shock phenomenon, with the back edges traveling
at a wave speed, and the tip of the rupture consisting in the place
where two shocks meet at a point.
\item The essential physical ingredients are dissipation and some toughening
that allows the back end of the rupture to retain its integrity. Static
hyperelasticity appears not to be relevant.
\item Predictions for rupture velocity are in satisfactory agreement with
experiment. Predictions for opening angle are less so, perhaps because
the computations have been performed in a triangular lattice, and
with a very simple form of dissipation.
\end{enumerate}
Additional physical question that have not yet been resolved are

\begin{enumerate}
\item Under what conditions do cracks in rubber creep, and when instead
are they supersonic?
\item What is the origin of rupture path oscillations reported by \citet{Deegan.02.rubber}?
\end{enumerate}
\begin{ack}
Jim Rice in some lengthy emails first pointed out the possibility
of trying to explain the rubber rupture experiments with the neo-Hookean
theory. Bertram Brogberg contributed similar comments not long after.
Leonid Slepyan pointed out that stress is singular in the continuum
version of the rupture theory. K Ravi-Chandar pressed me to expand
on many points, particularly the conventional theory of finite elasticity.
Paul Petersan, Robert Deegan, and Harry Swinney provided all the experimental
data in the paper, and took part in many discussions over how to explain
them. Thanks to the National Science Foundation for support from DMR-0401766
and DMR-0101030.
\end{ack}
\appendix

\section{Reduction to 2 dimensions\label{sec:Reduction-to-2}}

This Appendix shows in two different ways how to obtain an effective
two-dimensional equation of motion for a rubber sheet. In the first,
method, the incompressibility of rubber is used to calculate the thickness
of the sheet at every point, thus expressing displacements across
the thickness of the sheet ($z$ direction) in terms of the extensions
along $x$ and $y$ (Figure \ref{cap:Illustration-of-reference}).

\subsection{First method}

Rubber is highly incompressible, so one can set\begin{equation}
\det\left|\frac{\partial\vec{u}}{\partial\vec{r}}\right|=0.\label{eq:det}\end{equation}
For a thin sheet, assume that one can neglect $\partial u^{y}/\partial z$
and $\partial u^{x}/\partial z$, on the grounds that $u^{x}$ and
$u^{y}$ should be uniform through the thickness of the sheet. Then
Eq. \prettyref{eq:det} becomes \begin{equation}
\frac{\partial u^{z}}{\partial r_{z}}\left(\frac{\partial u^{x}}{\partial r_{x}}\frac{\partial u^{y}}{\partial r_{y}}-\frac{\partial u^{x}}{\partial r_{y}}\frac{\partial u^{y}}{\partial r_{x}}\right)=1.\label{eq:incompressibility}\end{equation}
Since $u^{x}$ and $u^{y}$ are assumed to be independent of $z$,
one can write\[
u^{z}=r_{z}\,\left(\frac{\partial u^{x}}{\partial r_{x}}\frac{\partial u^{y}}{\partial r_{y}}-\frac{\partial u^{x}}{\partial r_{y}}\frac{\partial u^{y}}{\partial r_{x}}\right)^{-1}\]
which is odd in $r_{z}$ In moving to a two--dimensional theory, replace
all quantities by their averages across the sheet. That is, if the
sheet has thickness $t$ in the reference frame, then for example,\begin{equation}
E_{xz}(r_{x},r_{y})\equiv\int_{-t/2}^{t/2}\frac{dr_{z}}{t}\, E_{xz}(r_{x},r_{y},r_{z}).\label{eq:average_thickness}\end{equation}
Consider now \[
E_{xz}(\vec{r})=\frac{1}{2}\left(\frac{\partial u^{x}}{\partial r_{x}}\frac{\partial u^{x}}{\partial r_{z}}+\frac{\partial u^{y}}{\partial r_{x}}\frac{\partial u^{y}}{\partial r_{z}}+\frac{\partial u^{z}}{\partial r_{x}}\frac{\partial u^{z}}{\partial r_{z}}\right).\]
To obtain the two--dimensional version of this quantity, note that
the first two terms vanish because of the derivatives with respect
to $r_{z}$ while the last term is odd in $r_{z}$, and vanishes when
averaged across the sheet thickness in \prettyref{eq:average_thickness}
. Therefore, in the two--dimensional theory, one can take $E_{xz}=E_{yz}=0.$
Finally, consider \begin{equation}
E_{zz}=\frac{1}{2}\left(\left[\frac{\partial u^{z}}{\partial r_{z}}\right]^{2}-1\right)=\frac{1}{2}\left(\left(\frac{\partial u^{x}}{\partial r_{x}}\frac{\partial u^{y}}{\partial r_{y}}-\frac{\partial u^{x}}{\partial r_{y}}\frac{\partial u^{y}}{\partial r_{x}}\right)^{-2}-1\right).\label{eq:AI5}\end{equation}
The derivatives appearing in the denominator of \prettyref{eq:AI5}
can be expressed in terms of the two--dimensional invariants in Eq.
\prettyref{eq:I1I2} as

\[
E_{zz}=\frac{1}{2}\left(\frac{1}{4I_{2}+2I_{1}+1}-1\right),\]
which is Eq. \prettyref{eq:Ezz}.

\subsection{Second Method}

Specialize to the case of a Neo--Hookean material. Note that raised
indices are employed on $u$ because elsewhere in the manuscript subscripts
are needed to index the locations of multiple particles; raised indices
are just ordinary Cartesian components of the vector $\vec{u}$. For
an incompressible solid the Cauchy stress tensor is \citep{Ogden1984}
\begin{equation}
T_{\alpha\beta}=\rho c^{2}\frac{\partial u^{\alpha}}{\partial r_{\gamma}}\frac{\partial u^{\beta}}{\partial r_{\gamma}}-p\delta_{\alpha\beta},\label{eq:Tab}\end{equation}
where $p$ is a pressure that must be determined by the condition
of incompressibility. To find an equation of motion, one needs the
Piola-Kirkhoff stress tensor, which for an incompressible material
takes the form\begin{equation}
S_{\alpha\beta}=\frac{\partial r_{\alpha}}{\partial u^{\lambda}}T_{\lambda\beta}=\rho c^{2}\frac{\partial u^{\beta}}{\partial r_{\alpha}}-p\frac{\partial r_{\alpha}}{\partial u^{\beta}}.\end{equation}
Writing out the equation of motion gives

\begin{equation}
\rho\frac{\partial^{2}u^{\alpha}}{\partial t^{2}}=\frac{\partial}{\partial r_{\lambda}}S_{\lambda\alpha}=\frac{\partial}{\partial r_{\lambda}}\left[\rho c^{2}\frac{\partial u^{\alpha}}{\partial r_{\lambda}}-p\frac{\partial r_{\lambda}}{\partial u^{\alpha}}\right].\end{equation}
From Eq. \prettyref{eq:incompressibility} one can write $\partial r_{\alpha}/\partial u^{\beta}$
as \begin{equation}
\left(\begin{array}{cc}
\frac{\partial r_{x}}{\partial u^{x}} & \frac{\partial r_{y}}{\partial u^{x}}\\
\frac{\partial r_{x}}{\partial u^{y}} & \frac{\partial r_{y}}{\partial u^{y}}\end{array}\right)=\left(\begin{array}{cc}
\frac{\partial u^{x}}{\partial r_{x}} & \frac{\partial u^{y}}{\partial r_{x}}\\
\frac{\partial u^{x}}{\partial r_{y}} & \frac{\partial u^{y}}{\partial r_{y}}\end{array}\right)^{-1}=\frac{\partial u^{z}}{\partial r_{z}}\left(\begin{array}{cc}
\frac{\partial u^{y}}{\partial r_{y}} & -\frac{\partial u^{y}}{\partial r_{x}\strut}\\
-\frac{\partial u^{x}}{\partial r_{y}} & \frac{\partial u^{x}}{\partial r_{x}}\end{array}\right).\end{equation}
 Now examine the equation of motion for $u^{x}$, 

\begin{equation}
\rho\frac{\partial^{2}u^{x}}{\partial t^{2}}=\left(\frac{\partial}{\partial r_{\lambda}}\rho c^{2}\frac{\partial u^{\alpha}}{\partial r_{\lambda}}\right)-\frac{\partial}{\partial r_{x}}\left(p\frac{\partial u^{z}}{\partial r_{z}}\frac{\partial u^{y}}{\partial r_{y}}\right)+\frac{\partial}{\partial r_{y}}\left(p\frac{\partial u^{z}}{\partial r_{z}}\frac{\partial u^{y}}{\partial r_{x}}\right).\label{eq:uxmot1}\end{equation}
Compare this result with the one that would come by inserting the
constraint at the outset: Using Eq. \prettyref{eq:AI5} one finds

\begin{eqnarray}
\rho\frac{\partial^{2}u^{x}}{\partial t^{2}} & = & -\frac{\delta}{\delta u^{x}(\vec{r)}}\int d\vec{r}\:\rho c^{2}(E_{xx}+E_{yy}+E_{zz})\nonumber \\
 & = & \left(\frac{\partial}{\partial r_{\lambda}}\rho c^{2}\frac{\partial u^{\alpha}}{\partial r_{\lambda}}\right)-\frac{\partial}{\partial r_{x}}\left[\rho c^{2}\left(\frac{\partial u^{z}}{\partial r_{z}}\right)^{3}\frac{\partial u^{y}}{\partial r_{y}}\right]\label{eq:uxmot2}\\
 &  & \quad\quad\quad\quad\quad\quad\:\:+\frac{\partial}{\partial r_{y}}\left[\rho c^{2}\left(\frac{\partial u^{z}}{\partial r_{z}}\right)^{3}\frac{\partial u^{y}}{\partial r_{x}}\right]\nonumber \end{eqnarray}
Eqs. \prettyref{eq:uxmot1} and \prettyref{eq:uxmot2} are the same
provided that\begin{equation}
p=\rho c^{2}\left(\frac{\partial u^{z}}{\partial r_{z}}\right)^{2}.\label{eq:pressure}\end{equation}
Thus the equation of motion obtained by employing the constraint in
Eq. \prettyref{eq:Ezz} is compatible with the equation of motion
one obtains from the Piola-Kirkhoff stress tensor so long as one uses
Eq. \prettyref{eq:pressure} for the pressure. Furthermore, this expression
for the pressure is precisely what is needed so that $T_{zz}$ vanishes
in Eq. \prettyref{eq:Tab}, and that in turn is what one would expect
as the appropriate boundary condition for a thin sheet.

\section{Sound Speeds\label{sec:Sound-Speeds}}

Given an energy functional \begin{eqnarray*}
U & = & \rho\int d\vec{r}\:\energy(I_{1},I_{2}),\end{eqnarray*}
where $\rho$ is the mass per area measured in the reference frame,
the aim of this appendix is to calculate sound speeds. The same results
are found for example in \citep[v. 1, pp. 120, 263]{Eringen.74},
but to obtain the simple expressions for longitudinal and shear waves
needed here, it may be easier to begin again than to work backwards
through so much notation. To begin, find how $U$ varies when there
is a small change in $u$: \[
\frac{1}{\rho}\frac{\delta U}{\delta u^{\gamma}(\vec{r})}=-\sum_{\alpha,\beta}\left[\frac{\partial^{2}u^{\gamma}}{\partial r_{\alpha}\partial r_{\beta}}\frac{\partial\energy}{\partial E_{\alpha\beta}}+\frac{1}{2}\left\{ \frac{\partial u^{\gamma}}{\partial r_{\alpha}}\frac{\partial}{\partial r_{\beta}}+\frac{\partial u^{\gamma}}{\partial r_{\beta}}\frac{\partial}{\partial r_{\alpha}}\right\} \frac{\partial\energy}{\partial E_{\alpha\beta}}\right]\]

Take $\partial e/\partial E_{\alpha\beta}$ to be symmetric under
interchange of $\alpha$ and $\beta$. This is only true if from now
on whenever one sees $E_{xy}$ in some term in the free energy, one
replaces it by $\left(E_{xy}+E_{yx}\right)$/2, and we will have to
be careful to do that. However, assuming this symmetry, one can write

\begin{eqnarray}
\frac{1}{\rho}\frac{\delta U}{\delta u^{\gamma}(\vec{r})} & = & -\sum_{\alpha,\beta}\left[\frac{\partial^{2}u^{\gamma}}{\partial r_{\alpha}\partial r_{\beta}}\frac{\partial\energy}{\partial E_{\alpha\beta}}+\sum_{\alpha'\beta'}\frac{\partial u^{\gamma}}{\partial r_{\alpha}}\frac{\partial E_{\alpha'\beta'}}{\partial r_{\beta}}\frac{\partial^{2}\energy}{\partial E_{\alpha\beta}\partial E_{\alpha'\beta'}}\right]\label{eq:dW1;a}\\
 &  & \!\!\!\!\!\!\!\ \!\!\!\!\!\!\!\ \!\!\!\!\!\!\!\ \!\!\!\!\!\!\!\ \!\!\!\!\!\!\!\ \!\!\!\!\!\!\!\ \!\!\!\!\!\!\!\ \!\!\!\!\!\!\!\ \!\!\!\!\!\!\!\ \!\!\!\!\!\!\!\ =-\sum_{\alpha,\beta}\left[\frac{\partial^{2}u^{\gamma}}{\partial r_{\alpha}\partial r_{\beta}}\frac{\partial\energy}{\partial E_{\alpha\beta}}+\sum_{\alpha'\beta'\gamma'}\frac{1}{2}\frac{\partial u^{\gamma}}{\partial r_{\alpha}}\left({\displaystyle \frac{\partial}{\partial r_{\beta}}\left[\frac{\partial u^{\gamma'}}{\partial r_{\alpha'}}\frac{\partial u^{\gamma'}}{\partial r_{\beta'}}\right]}\right)\frac{\partial^{2}\energy}{\partial E_{\alpha\beta}\partial E_{\alpha'\beta'}}\right]\nonumber \\
 &  & =-\sum_{\alpha,\beta}\left[\frac{\partial^{2}u^{\gamma}}{\partial r_{\alpha}\partial r_{\beta}}\frac{\partial\energy}{\partial E_{\alpha\beta}}+\sum_{\alpha'\beta'\gamma'}\frac{\partial u^{\gamma}}{\partial r_{\alpha}}\frac{\partial u^{\gamma'}}{\partial r_{\alpha'}}\frac{\partial^{2}u^{\gamma'}}{\partial r_{\beta'}\partial r_{\beta}}\frac{\partial^{2}\energy}{\partial E_{\alpha\beta}\partial E_{\alpha'\beta'}}\right]\nonumber \end{eqnarray}

Now take $u$ to represent a sheet loaded up in biaxial strain, and
superpose a small amplitude wave with polarization $\vec{\epsilon}$
traveling with wave vector $\vec{k}$. Take $\lambda_{\gamma}$to
be the extension factor along direction $\gamma$, and $r_{\gamma}$
to be a position coordinate in the material frame. Keep only terms
of order $\epsilon$. Inserting such a plane wave into Eq. \prettyref{eq:dW1;a},
the result is

\begin{eqnarray}
\frac{1}{\rho}\frac{\delta U}{\delta u^{\gamma}(\vec{r)}} & \approx & \sum_{\alpha,\beta}\left[k_{\alpha}k_{\beta}\epsilon_{\gamma}\frac{\partial\energy}{\partial E_{\alpha\beta}}+\sum_{\alpha'\beta'\gamma'}\lambda_{\gamma}\delta_{\gamma\alpha}\:\lambda_{\gamma'}\delta_{\gamma'\alpha'}k_{\beta'}k_{\beta}\epsilon_{\gamma'}\frac{\partial^{2}\energy}{\partial E_{\alpha\beta}\partial E_{\alpha'\beta'}}\right]\nonumber \\
 & = & \sum_{\alpha,\beta}\left[k_{\alpha}k_{\beta}\epsilon_{\gamma}\frac{\partial\energy}{\partial E_{\alpha\beta}}\right]+\left[\sum_{{{\beta\atop \beta'\gamma}}}\lambda_{\gamma}\:\lambda_{\gamma'}k_{\beta'}k_{\beta}\epsilon_{\gamma'}\frac{\partial^{2}\energy}{\partial E_{\gamma\beta}\partial E_{\gamma'\beta'}}\right]\nonumber \\
 & = & \sum_{\beta\beta'\gamma'}\left[k_{\beta}k_{\beta'}\epsilon_{\gamma}\frac{\partial\energy}{\partial E_{\beta\beta'}}+\lambda_{\gamma}\:\lambda_{\gamma'}k_{\beta'}k_{\beta}\epsilon_{\gamma'}\frac{\partial^{2}\energy}{\partial E_{\gamma\beta}\partial E_{\gamma'\beta'}}\right]\nonumber \\
 & = & \sum_{\beta\beta'\gamma'}\left[\delta_{\gamma\gamma'}\frac{\partial\energy}{\partial E_{\beta\beta'}}+\lambda_{\gamma}\lambda_{\gamma'}\frac{\partial^{2}\energy}{\partial E_{\gamma\beta}\partial E_{\gamma'\beta'}}\right]k_{\beta'}k_{\beta}\epsilon_{\gamma'}.\end{eqnarray}

Therefore, one has an equation of motion for sound waves

\begin{eqnarray}
\rho\omega^{2}\epsilon_{\gamma} & = & \frac{\delta U}{\delta u^{\gamma}(\vec{r})}\nonumber \\
\Rightarrow\omega^{2}\epsilon_{\gamma} & = & \sum_{\gamma'\beta\beta'}\left[\delta_{\gamma\gamma'}\frac{\partial\energy}{\partial E_{\beta\beta'}}+\lambda_{\gamma}\lambda_{\gamma'}\frac{\partial^{2}\energy}{\partial E_{\gamma\beta}\partial E_{\gamma'\beta'}}\right]k_{\beta}k_{\beta'}\epsilon_{\gamma'}.\label{eq:dW4}\end{eqnarray}

\subsection{Specific expressions for longitudinal and shear waves}

Take $\vec{k}=k(1,0)$. Assume that \begin{equation}
\frac{\partial^{2}\energy}{\partial E_{xx}\partial E_{xy}}=\frac{\partial^{2}\energy}{\partial E_{yy}\partial E_{xy}}=0.\end{equation}
This will always be the case in biaxial strain just so long as the
energy only depends upon strain through the combinations in $I_{1}$
and $I_{2}$. Longitudinal waves are found by looking for a wave polarized
along $x$, which means that only $\epsilon_{x}$ is nonzero, so $\gamma=\gamma'=x$.
Note in addition that in Eq. \prettyref{eq:dW4} one can have $k_{\beta}=k_{\beta'}=1$
only if $\beta=\beta'=x$. Therefore\begin{equation}
c_{lx}^{2}=\frac{\partial\energy}{\partial E_{xx}}+\lambda_{x}^{2}\frac{\partial^{2}\energy}{\partial E_{xx}^{2}}.\label{eq:longitudinal0}\end{equation}

Next look for a wave polarized along $y.$ Now $\gamma=\gamma'=y$.
One still has to have $\beta=\beta'=x$. Therefore for shear waves

\begin{equation}
c_{sx}^{2}=\frac{\partial\energy}{\partial E_{xx}}+\lambda_{y}^{2}\frac{\partial^{2}\energy}{\partial E_{xy}^{2}}\:\mbox{Assuming symmetry}\label{eq:shear0}\end{equation}

It is easy to use Eq. \ref{eq:shear0} improperly. It is only valid
if $e$ is treated as a symmetrical function of $E_{xy}$ and $E_{yx}$,
and if partial derivatives with respect to these two quantities are
independent. It is hard to remember to retain this convention, and
it is safer simply to set $E_{xy}=E_{yx}$ and treat $e$ just as
a function of one of them. In this case, one must write

\begin{equation}
c_{sx}^{2}=\frac{\partial\energy}{\partial E_{xx}}+\frac{\lambda_{y}^{2}}{4}\frac{\partial^{2}\energy}{\partial E_{xy}^{2}}.\label{eq:shear0.1}\end{equation}
The analogous expressions for wave speeds along $y$ follow by flipping
the roles of $x$ and $y$.

The expressions for sound speeds take much simpler forms if one establishes
spatially uniform states in the rubber and considers small uniform
disturbances. Return to Eqs. \prettyref{eq:longitudinal} and \prettyref{eq:shear}.
Establish the displacement field 

\begin{equation}
\vec{u}=(\lambda_{x}x+s_{xy}y,\lambda_{y}y+s_{yx}x).\label{eq:deformation_b}\end{equation}

Specialize now to the case where a sample is subject to uniform bi--axial
strain, and sheared in the $y$ direction, so that $s_{xy}$ is zero.
Then inserting Eq. \prettyref{eq:deformation_b} into Eq. \ref{eq:Eab}
gives\begin{eqnarray}
E_{xx} & = & {\textstyle \frac{1}{2}}\left(\lambda_{x}^{2}+s_{yx}^{2}-1\right)\\
E_{yy} & = & {\textstyle \frac{1}{2}}\left(\lambda_{y}^{2}-1\right)\\
E_{xy} & = & {\textstyle E_{yx}=\frac{1}{2}}\lambda_{y}s_{yx}\end{eqnarray}

Derivatives with respect to components of the strain tensor can all
now be expressed in terms of the new variables $\lambda_{x}$, $\lambda_{y}$,
and $s_{yx}$. Since these variables correspond exactly to quantities
one controls experimentally, it is good to express sound speeds in
terms of them. One has

\begin{equation}
\frac{\partial\left\{ E_{xx}E_{yy}E_{xy}\right\} }{\partial\left\{ \lambda_{x}\lambda_{y}s_{yx}\right\} }=\left(\begin{array}{ccc}
\lambda_{x} & 0 & s_{yx}\\
0 & \lambda_{y} & 0\\
0 & s_{yx}/2 & \lambda_{y}/2\end{array}\right).\end{equation}

Inverting this matrix, one has 

\begin{equation}
\frac{\partial\left\{ \lambda_{x}\lambda_{y}s_{yx}\right\} }{\partial\left\{ e_{xx}e_{yy}e_{xy}\right\} }=\left(\begin{array}{ccc}
\frac{1}{\lambda_{x}} & \frac{s_{yx}^{2}}{\lambda_{x}\lambda_{y}^{2}} & -\frac{2s_{yx}}{\lambda_{x}\lambda_{y}}\\
0 & \frac{1}{\lambda_{y}} & 0\\
0 & -\frac{s_{yx}}{\lambda_{y}^{2}} & \frac{2}{\lambda_{y}}\end{array}\right).\end{equation}
 Therefore, one can write\begin{eqnarray}
\frac{\partial}{\partial E_{xx}} & = & \frac{1}{\lambda_{x}}\frac{\partial}{\partial\lambda_{x}}\label{eq:exx_l}\\
\frac{\partial}{\partial E_{xy}} & = & -\frac{2s_{yx}}{\lambda_{y}\lambda_{y}}\frac{\partial}{\partial\lambda_{x}}+\frac{2}{\lambda_{y}}\frac{\partial}{\partial s_{yx}}.\label{eq:exy_l}\end{eqnarray}

Inserting Eq. \prettyref{eq:exx_l} into Eq. \prettyref{eq:longitudinal}
and Eq. \prettyref{eq:exy_l} into Eq. \prettyref{eq:shear} and evaluating
at $s_{yx}=0$ gives Eqs. \prettyref{eq:soundspeeds}.

\section{Force computation\label{sec:Force-computation}}

I record here some methods used to calculate forces in molecular dynamics
that assist in writing computer code, and that may not have been published
previously. The force on component $\alpha$ of particle $l$ is defined
to be\begin{equation}
F_{l}^{\alpha}\equiv-\frac{\partial U}{\partial u_{l}^{\alpha}}.\end{equation}
For the purposes of writing computer code, it is not efficient to
proceed directly with this expression, because in the course of computing,
say, $F_{1}$ one might calculate some quantities that will also appear
in $F_{2}$ and efficient code will duplicate as little computation
as possible. Therefore, define the \emph{insertion operator $\mathcal{I}_{l}$.}
Any term that is multiplied by \emph{$\mathcal{I}_{l}$} is to be
inserted into the memory location that holds the force on particle
$l$. So one computes

\begin{equation}
\sum_{l}F_{l}^{\alpha}I_{l}/m=-\sum_{li}\mathcal{I}_{l}\left[\frac{\partial w}{\partial I_{1}^{i}}\frac{\partial I_{1}^{i}}{\partial u_{l}^{\alpha}}+\frac{\partial w}{\partial I_{2}^{i}}\frac{\partial I_{2}^{i}}{\partial u_{l}^{\alpha}}\right]\end{equation}

Thus one must compute a sum of two terms. The first is

\begin{equation}
\sum_{li,j\in n(i)}\mathcal{I}_{l}\frac{\partial w}{\partial I_{1}^{i}}\frac{1}{3\lspace^{2}}(\delta_{il}-\delta_{jl})u_{ij}^{\alpha}=\sum_{i,j\in n(i)}(\mathcal{I}_{i}-\mathcal{I}_{j})\frac{\partial w}{\partial I_{1}^{i}}\frac{1}{3\lspace^{2}}u_{ij}^{\alpha}\end{equation}
(but this term vanishes if $u_{ij}>\lspace\lambda_{f}$)

The second is\begin{equation}
\sum_{ij\in n(i)}(\mathcal{I}_{i}-\mathcal{I}_{j})u_{ij}^{\alpha}\frac{\partial w}{\partial J_{2}^{i}}\frac{3}{4\lspace^{4}}\left[\frac{2}{3}F_{i}-\frac{4}{9}(\vec{u}_{ij}\cdot\vec{u}_{ij}-\lspace^{2})\right],\end{equation}
(but vanish if $u_{ij}>\lspace\lambda_{f})$ or, if one takes the
alternate representation of $I_{2}^{i},$\begin{equation}
\sum_{ij\in n(i)}(\mathcal{I}_{i}-\mathcal{I}_{j})u_{ij}^{\alpha}\frac{\partial w}{\partial I_{2}^{i}}\frac{9}{8\lspace^{4}}\left[\frac{4}{9}(\vec{u}_{ij}\cdot\vec{u}_{ij}-\lspace^{2})\right]+\sum_{l}\mathcal{I}_{l}\frac{\partial w}{\partial I_{2}^{i}}\frac{9}{8\lspace^{4}}\frac{\partial H_{i}}{\partial u_{l}^{\alpha}},\end{equation}
The final term to compute is (with $g_{ijk}\equiv(\vec{u}_{ij}\cdot\vec{u}_{ik}+2a^{2})^{2})$\begin{equation}
H_{i}\equiv\frac{1}{27}\sum_{j\neq k\in n(i)}(\vec{u}_{ij}\cdot\vec{u}_{ik}+2\lspace^{2})^{2}h(u_{ij})h(u_{ik})\equiv\frac{1}{27}\sum_{j\neq k\in n(i)}g_{ijk}h_{ij}h_{ik}\end{equation}
 and derivatives of this object contribute to the force \begin{equation}
-\sum_{{{i\atop j\neq k\in n(i)}}}\frac{\partial w}{\partial J_{2}^{i}}\frac{1}{27}\frac{9}{8\lspace^{4}}\left\{ \begin{array}{cc}
 & h'_{ij}h_{ik}g_{ijk}\frac{u_{ij}^{\alpha}}{u_{ij}}\left(\mathcal{I}_{i}-\mathcal{I}_{j}\right)+h'_{ik}h_{ij}g_{ijk}\frac{u_{ik}^{\alpha}}{u_{ik}}\left(\mathcal{I}_{i}-\mathcal{I}_{k}\right)\\
\noalign{\vskip0in}\\
+ & h{}_{ij}h_{ik}g'_{ijk}\left(\mathcal{I}_{i}\left\{ u_{ij}^{\alpha}+u_{ik}^{\alpha}\right\} -\mathcal{I}_{k}u_{ij}^{\alpha}-\mathcal{I}_{j}u_{ik}^{\alpha}\right)\end{array}\right\} .\end{equation}
Note in all these expression that various quantities need only be
computed once and inserted into registers for particles $i$, $j,$
and $k.$ The insertion operators $\mathcal{I}_{i}$ , $\mathcal{I}_{j}$
, and $\mathcal{I}_{k}$ keep track of which quantities to put where.

\section{Lattice Instabilities}

\begin{figure}[h]
\begin{center}\includegraphics[%
  width=1.0\columnwidth,
  keepaspectratio]{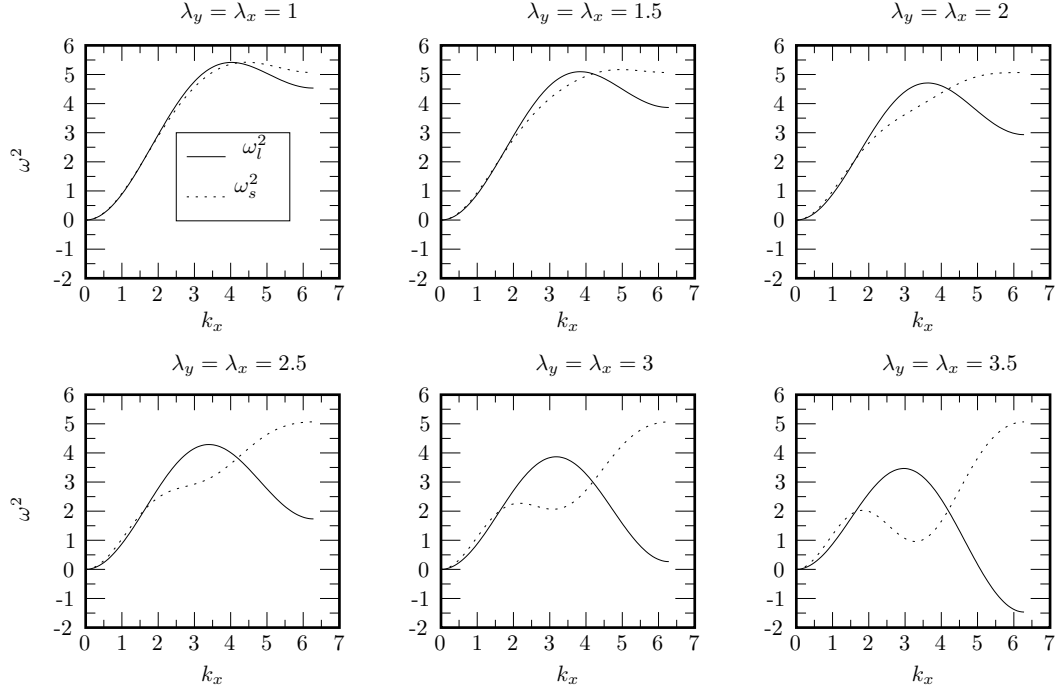}\end{center}

\caption{Frequency squared as a function of wave number for various levels
of uniform biaxial strain, using the energy potential function in
Eq. \prettyref{eq:J2ia}. As the strain increases, the lattice becomes
unstable to a distortion in which every other column of atoms moves
in opposite directions, similar to motions of atoms in optical modes,
or the Peierls distortion. \label{cap:Longitudinal-and-shear}}
\end{figure}

\begin{figure}[H]
\begin{center}\includegraphics[%
  width=1.0\columnwidth,
  keepaspectratio]{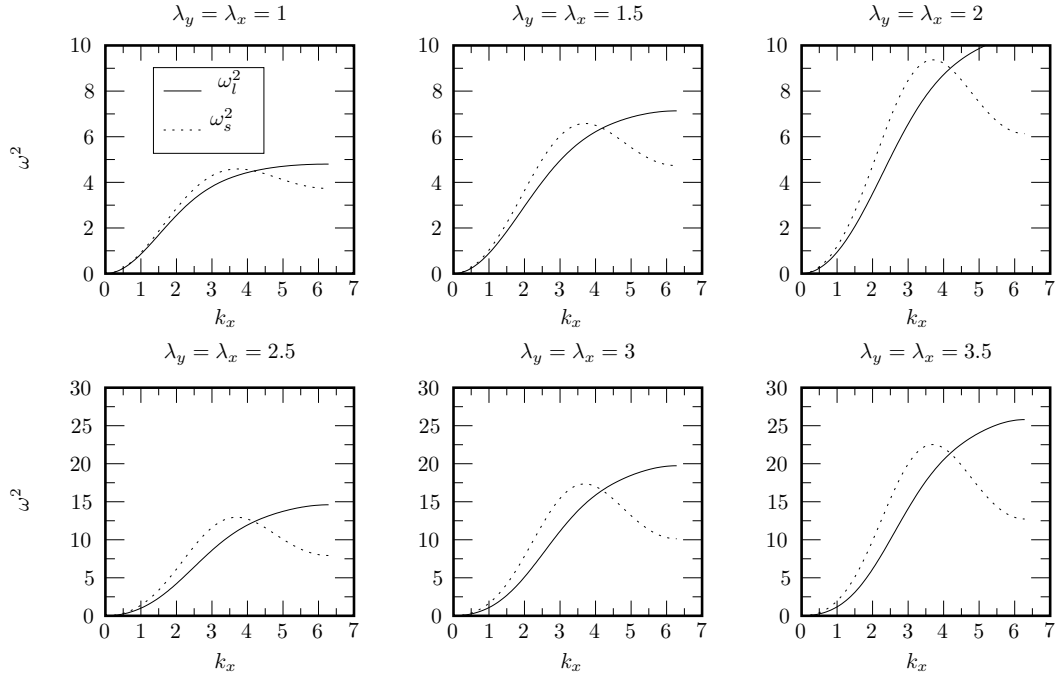}\end{center}

\caption{Frequency squared as a function of wave number for various levels
of uniform biaxial strain, using the energy potential function in
Eq. \prettyref{eq:J2ib}. With this representation of the strain invariant,
the uniform lattice remains stable against high-frequency distortions
over the range of extensions employed experimentally. \label{cap:Longitudinal-and-shear2}}
\end{figure}

I found numerically that when I employed Eq. \prettyref{eq:J2ia},
the uniformly strained lattice would spontaneously develop a striped
pattern when stretched beyond a critical value. To analyze this problem,
I computed the phonon dynamical matrix \citep[Eq. 13.8, p. 307]{Marder.00.CMP}.
The calculation is a straightforward exercise in phonon physics, and
no details need to be reported. Negative eigenvalues of this matrix
indicate instability of the uniform state. As shown in Figure \ref{cap:Longitudinal-and-shear},
for uniform biaxial strain a bit above $\lambda_{x}=\lambda_{y}=3,$
the spatially uniform lattice becomes unstable. However, if one employs
instead Eq. \prettyref{eq:J2ib}, then as shown in Fig. \ref{cap:Longitudinal-and-shear2},
the lattice remains stable. For this reason, Eq. \prettyref{eq:J2ib}
was usually employed, despite its greater numerical cost. It would
be interesting to see whether the instability in Fig. \ref{cap:Longitudinal-and-shear}
is related to strain crystallization.


\section{Solution of Discrete System\label{app:discrete}}

This Appendix contains the steady-state solution of \begin{equation}
\ddot{u}_{i}^{\alpha}=\frac{2c^{2}}{3\lspace^{2}}\sum_{j\in n(i)}\left(u_{ij}^{\alpha}+\beta\dot{u}_{ij}^{\alpha}\right)\theta(\lambda_{f}-u_{ij}).\end{equation}
The methods employed are those of \citet{Slepyan.81,Slepyan.82,Slepyan.02};
see also \citep{Marder.95.JMPS,Marder.IJF.04}. \citet[p. 478]{Slepyan.02}
notes the existence of supersonic solutions for a related problem.
However, the steps below are worth recording in detail because the
particular combination of Kelvin viscosity, Mode III, and a strip
of finite height needed here has not been published, although no especially
new ideas are involved.

Move to notation that explicitly describes locations in a triangular
lattice, $u(m,n),$ where $u$ describes the vertical motion of atoms
only, since the horizontal motion is neglected, and where $m\in(-\infty\dots-3,-2,-1,0,1,2,3,\dots\infty)$
and $n\in(\dots-5/2,\ -3/2,\ -1/2,\ 1/2,\ \ 3/2,\ 5/2\dots$. Then
in steady state, one has the symmetries\begin{eqnarray}
u(m,n,t) & = & u(m+1,n,t+\lspace/v)\label{eq:TL3;a}\\
u(m,n,t) & = & -u(m,-n,t-\lspace[1/2-g_{n}]/v)\label{eq:TL3;b}\\
u(m,1/2,t) & = & -u(m,-1/2,t-\lspace/2v),\label{eq:TL3;c}\end{eqnarray}
 where\begin{eqnarray}
g_{n} & = & \begin{cases}
1 & \textrm{if }n=3/2,\ 7/2...\\
0 & ,\textrm{if }n=1/2,\ 5/2...\end{cases}\label{TL1}\end{eqnarray}

Assuming that a crack is in steady state, we can therefore eliminate
the variable $m$ entirely from the equation of motion, by defining
\begin{equation}
u_{n}(t)=u(0,n,t).\label{93a0}\end{equation}
 Next, define dimensionless variables\begin{equation}
\tilde{t}=tc/\lspace;\ \tilde{\beta}=\beta c/\lspace,\quad\mbox{and}\ \tilde{v}=v/c.\label{eq:dimensionless}\end{equation}
Then \begin{widetext}\begin{eqnarray*}
\ddot{u}_{n}(\tilde{t}) & = & \frac{2}{3}(1+\tilde{\beta}\frac{\partial}{\partial\tilde{t}})\left[\begin{array}{lcl}
+u_{n+1}(\tilde{t}-(g_{n+1}-1)/\tilde{v}) &  & +u_{n+1}(\tilde{t}-g_{n+1}/\tilde{v})\\
+u_{n}(\tilde{t}+1/\tilde{v}) & -6u_{n}(\tilde{t}) & +u_{n}(\tilde{t}-1/\tilde{v})\\
+u_{n-1}(\tilde{t}-(g_{n-1}-1)/\tilde{v}) &  & +u_{n-1}(\tilde{t}-g_{n-1}/\tilde{v})\end{array}\right]\\
 &  & \textrm{if $n>1/2$ and }\end{eqnarray*}
 \begin{equation}
\ddot{u}_{1/2}(\tilde{t})=\frac{2}{3}\left[\begin{array}{lcl}
+(1+\tilde{\beta}\frac{\partial}{\partial\tilde{t}})u_{3/2}(\tilde{t})+(1+\tilde{\beta}\frac{\partial}{\partial\tilde{t}})u_{3/2}(\tilde{t}-1/\tilde{v})\\
+(1+\tilde{\beta}\frac{\partial}{\partial\tilde{t}})\left[u_{1/2}(\tilde{t}+1/\tilde{v})-4u_{1/2}(\tilde{t})+u_{1/2}(\tilde{t}-1/\tilde{v})\right]\\
+\theta(-\tilde{t})(1+\tilde{\beta}\frac{\partial}{\partial\tilde{t}})[u_{-1/2}(\tilde{t})-u_{1/2}(\tilde{t})]\\
+\theta(1/(2\tilde{v})-\tilde{t})(1+\tilde{\beta}\frac{\partial}{\partial\tilde{t}})[u_{-1/2}(\tilde{t}-1/\tilde{v})-u_{1/2}(\tilde{t})]\end{array}\right]\label{eq:TL3}\end{equation}
if $n=1/2$.

The time at which the bond between $u(0,1/2,\tilde{t})$ and $u(0,-1/2,\tilde{t})$
breaks has been chosen to be $t=0$, so that by symmetry the time
the bond between $u(0,1/2,\tilde{t})$ and $u(1,-1/2,\tilde{t})$
breaks is $1/2\tilde{v}$.

Above the crack line, the equations of motion are completely linear,
so it is simple to find the motion of every atom with $n>1/2$ in
terms of the behavior of an atom with $n=1/2$. Fourier transforming
in time gives \begin{equation}
-\omega^{2}u_{n}(\omega)=\frac{2}{3}(1-i\tilde{\beta}\omega)\left[\begin{array}{lll}
 & u_{n+1}(\omega) & [e^{i\omega(g_{n+1}-1)/\tilde{v}}+e^{i\omega(g_{n+1})/\tilde{v}}]\\
+ & u_{n}(\omega) & [e^{i\omega/\tilde{v}}-6+e^{-i\omega/\tilde{v}}]\\
+ & u_{n-1}(\omega) & [e^{i\omega(g_{n-1}-1)/\tilde{v}}+e^{i\omega(g_{n-1})/\tilde{v}}]\end{array}.\right]\label{eq:TL4}\end{equation}

Let \begin{equation}
u_{n}(\omega)=u_{1/2}(\omega)e^{k(n-1/2)-i\omega g_{n}/(2\tilde{v})}.\label{eq:TL5}\end{equation}
 Substituting Eq. \prettyref{eq:TL5} into Eq. \prettyref{eq:TL4},
and noticing that $g_{n}+g_{n+1}=1$ gives \begin{equation}
-\omega^{2}u_{1/2}(\omega)=(1-i\tilde{\beta}\omega)\frac{2}{3}\left[\begin{array}{lll}
 & u_{1/2}(\omega)e^{k} & [e^{i\omega(g_{n+1}+g_{n-2})/(2\tilde{v})}+e^{i\omega(g_{n+1}+g_{n})/(2\tilde{v})}]\\
+ & u_{1/2}(\omega) & [e^{i\omega/\tilde{v}}-6+e^{-i\omega/\tilde{v}}]\\
+ & u_{1/2}(\omega)e^{-k} & [e^{i\omega(g_{n-1}+g_{n-2})/(2\tilde{v})}+e^{i\omega(g_{n-1}+g_{n})/(2\tilde{v})}]\end{array}\right]\label{TL6}\end{equation}
\begin{equation}
\Rightarrow\frac{\omega^{2}}{1-i\tilde{\beta}\omega}+\frac{4}{3}\left[2\cosh(k)\cos(\omega/(2\tilde{v}))+\cos(\omega/\tilde{v})-3\right]=0.\label{TL7}\end{equation}
 \end{widetext}Defining \begin{equation}
\zeta=\frac{3-\cos(\omega/\tilde{v})-3\omega^{2}/[4(1-i\tilde{\beta}\omega)]}{2\cos(\omega/2\tilde{v})}\label{TL8.0}\end{equation}
 one has equivalently that \begin{equation}
\phi=e^{k}=\zeta+\sqrt{\zeta^{2}-1}\ \mbox{\ \  with }\quad\textrm{abs}(\phi)>1.\label{TL9.0}\end{equation}
 One can construct a solution which meets all the boundary conditions
by writing \begin{eqnarray}
u_{n}(\omega) & = & u_{1/2}(\omega)e^{-i\omega g_{n}/2\tilde{v}}[\frac{\phi^{[N+1/2-n]}-\phi^{-[N+1/2-n]}}{\phi^{N}-\phi^{-N}}]\nonumber \\
 & + & \frac{U_{N}(n-1/2)}{N}\frac{2\epsilon}{\epsilon^{2}+\omega^{2}}.\label{eq:TL11}\end{eqnarray}
 This solution equals $u_{1/2}$ for $n=1/2$, and equals $U_{N}2\epsilon/(\epsilon^{2}+\omega^{2})$
for $n=N+1/2$. The reason to introduce $\epsilon$ is that for $n=N+1/2$,
$u(m,n,t)=U_{N}$. The Fourier transform of this boundary condition
is a delta function, and hard to work with formally. To resolve uncertainties,
it is better to use instead the boundary condition \begin{equation}
u_{N+1/2}(t)=U_{N}e^{-\epsilon|t|},\label{TL11.1}\end{equation}
 and send $\epsilon$ to zero the end of the calculation. In what
follows, frequent use will be made of the fact that $\epsilon$ is
small.

The most interesting variable is not $u_{1/2}$, but the distance
between the bonds which will actually snap. Furthermore, the quantity
multiplied by the $\theta$ function in Eq. \ref{eq:TL3} is operated
upon by $(1+\beta\partial/\partial t)$ since dissipation stops operating
when bonds break. For this reason define \begin{equation}
U(t)={\frac{u_{1/2}(t)-u_{-1/2}(t)}{2}}={\frac{u_{1/2}(t)+u_{1/2}(t+1/2v)}{2}}.\label{eq:TL12}\end{equation}
\[
W(\tilde{t})=(1+\tilde{\beta}\frac{\partial}{\partial\tilde{t}})U(\tilde{t})\]
 Rewrite \ref{eq:TL3} as \begin{equation}
\ddot{u}_{1/2}(\tilde{t})=\frac{2}{3}\left[\begin{array}{lr}
+(1+\tilde{\beta}\frac{\partial}{\partial\tilde{t}})u_{3/2}(\tilde{t})+(1+\tilde{\beta}\frac{\partial}{\partial t})u_{3/2}(\tilde{t}-1/\tilde{v})\\
\noalign{\vskip-.1in}\\
+(1+\tilde{\beta}\frac{\partial}{\partial\tilde{t}})\left(u_{1/2}(\tilde{t}+1/\tilde{v})-4+u_{1/2}(\tilde{t})+u_{1/2}(\tilde{t}-1/\tilde{v})\right)\\
\noalign{\vskip-.1in}\\
-2U(\tilde{t})\theta(-\tilde{t})-2U(\tilde{t}-1/2\tilde{v})\theta(1/(2\tilde{v})-\tilde{t})\end{array}\right].\label{TL13}\end{equation}

Fourier transforming this expression using Eq. \prettyref{eq:TL11}
and defining \begin{equation}
U^{\pm}(\omega)=\int_{-\infty}^{\infty}d\tilde{t}\, e^{i\omega\tilde{t}}U(\tilde{t})\theta(\pm\tilde{t}),\label{L93.5.1a}\end{equation}
\begin{equation}
W^{\pm}(\omega)=\int_{-\infty}^{\infty}d\tilde{t\,}e^{i\omega\tilde{t}}W(\tilde{t})\theta(\pm\tilde{t}),\label{eq:L93.5.1aa}\end{equation}
 now gives \begin{equation}
(1-i\omega\tilde{\beta})u_{1/2}(\omega)F(\omega)-(1+e^{i\omega/2\tilde{v}})U^{-}(\omega)=-\frac{U_{N}}{N}\frac{2\epsilon}{\omega^{2}+\epsilon^{2}},\label{TL14.0}\end{equation}
 with \begin{equation}
F(\omega)=\left\{ \frac{\phi^{[N-1]}-\phi^{-[N-1]}}{\phi^{N}-\phi^{-N}}-2\zeta\right\} \cos(\omega/2\tilde{v})+1\label{L93.5.1b}\end{equation}
 Next, use Eq. \prettyref{eq:TL12}in the form \begin{equation}
W(\omega)=(1-i\omega\tilde{\beta}){\frac{(1+e^{-i\omega/2\tilde{v}})}{2}}u_{1/2}(\omega)\label{TL15}\end{equation}
 to obtain \begin{equation}
W(\omega)F(\omega)-2(\cos^{2}\omega/4\tilde{v})U^{-}(\omega)=-\frac{U_{N}}{N}\frac{2\epsilon}{\omega^{2}+\epsilon^{2}}.\label{TL14}\end{equation}
 Writing \begin{equation}
W(\omega)=W^{+}(\omega)+W^{-}(\omega)\label{L93a1}\end{equation}
 finally gives \begin{equation}
W^{+}(\omega)Q(\omega)+W^{-}(\omega)={U_{N}Q_{0}}\left[{\frac{1}{\epsilon+i\omega}}+{\frac{1}{\epsilon-i\omega}}\right],\label{eq:L93.5}\end{equation}
 with 

\begin{eqnarray}
Q & = & F/(F-1-\cos(\omega/2\tilde{v})).\end{eqnarray}
The Wiener-Hopf technique \citep{Noble.58} directs one to write \begin{equation}
Q(\omega)={\frac{Q^{-}(\omega)}{Q^{+}(\omega)}},\label{OD18}\end{equation}
 where $Q^{-}$ is free of poles and zeroes in the lower complex $\omega$
plane and $Q^{+}$ is free of poles and zeroes in the upper complex
plane. One can carry out this decomposition with the explicit formula
\begin{eqnarray}
Q^{\pm}(\omega) & = & \exp[\lim_{\epsilon\rightarrow0}\int{\frac{d\omega^{\prime}}{2\pi}}{\frac{\ln Q(\omega^{\prime})}{i\omega\mp\epsilon-i\omega^{\prime}}}].\label{eq:L93a.2;b}\end{eqnarray}

Separate Eq. \prettyref{eq:L93.5} into two pieces, one of which has
poles only in the lower half plane, and one of which has poles only
in the upper half plane: \begin{equation}
\frac{W^{+}(\omega)}{Q^{+}(\omega)}-\frac{Q_{0}U_{N}}{Q(0)}\frac{1}{(-i\omega+\epsilon)}=\frac{Q_{0}U_{N}}{Q^{-}(0)}\frac{1}{(i\omega+\epsilon)}-\frac{W^{-}(\omega)}{Q^{-}(\omega)}.\label{eq:L93.5a}\end{equation}
 Because the right and left hand sides of this equation have poles
in opposite sections of the complex plane, they must separately equal
a constant, ${\mathcal{C}}$. The constant must vanish, or $U^{-}$
and $U^{+}$ will behave as a delta function near $t=0$. So \begin{eqnarray}
W^{-}(\omega)=U_{N}\frac{Q_{0}Q^{-}(\omega)}{Q^{-}(0)(\epsilon+i\omega)}, & \textrm{and}\  & W^{+}(\omega)=U_{N}\frac{Q_{0}Q^{+}(\omega)}{Q^{-}(0)(\epsilon-i\omega)}.\label{eq:L93.6}\end{eqnarray}
 One now has an explicit solution for $W(\omega)$. Numerical evaluation
of $W(t)$ from Eq. \prettyref{eq:L93.6} is fairly straightforward,
using fast Fourier transforms. The most interesting quantity to obtain
is the separation between bonds opposite the crack line at $t=0,$
since by setting this quantity so that the bond snaps, one obtains
a consistent equation of motion. So one wants to find $U(0).$ To
obtain it, write\begin{equation}
U(\omega)=\frac{W^{+}(\omega)+W^{-}(\omega)}{1-i\tilde{\beta}\omega}.\label{eq:L93.7}\end{equation}
 The denominator of Eq. \prettyref{eq:L93.7} has a pole in the lower
half plane at \begin{equation}
\omega=-i/\tilde{\beta}\equiv-i\omega_{0}\label{eq:L93.8}\end{equation}
and this pole must be subtracted off in order to form $U^{-}.$ So
one has\begin{equation}
U^{-}(\omega)=\frac{W^{-}(\omega)-W^{-}(-i\omega_{0})}{1-i\tilde{\beta}\omega}.\label{eq:L93.9}\end{equation}
In order to find $U(t=0)$ it is sufficient to find\[
U(t=0)=\lim_{\omega\rightarrow\infty}i\omega U^{-}(\omega).\]
The reason is that $U^{-}(t)$ is zero for all positive $t,$ dropping
to zero right at $t=0.$ The value of $U(t=0)$ is given by the discontinuity
in $U^{-}(t).$ However, if $U^{-}(\omega)$ decays faster than $1/\omega$
for large $\omega,$ then the inverse Fourier transform of $U^{-}(\omega)$
must be continuous at $t=0.$ Therefore, from the coefficient of $1/i\omega,$
one can pick out the value of $U(t=0).$ Since $W^{-}(t)$ like $U^{-}(t)$
has a step-function discontinuity at $t=0,$ $W^{-}(\omega)$ decays
as $1/i\omega$ as $\omega$ goes to infinity. Thus one deduces immediately
from \prettyref{eq:L93.9} that \begin{equation}
U(t=0)=\omega_{0}W^{-}(-i\omega_{0}).\label{eq:L93.10}\end{equation}
Returning to \prettyref{eq:L93.6} one has

\begin{equation}
U_{0}\equiv U(t=0)=U_{N}\frac{Q_{0}Q^{-}(-i\omega_{0})}{Q^{-}(0)}.\label{eq:L93.11}\end{equation}
Note that the height of the top of the system $U_{N}$ obeys\[
\frac{\sqrt{3}}{2}\lspace(N+1/2)\lambda_{y}=U_{N},\]
and that $Q_{0}=Q(0)=1/(2N+1)$ so \[
U_{0}=\frac{\sqrt{3}}{4}\lspace\lambda_{y}\frac{Q^{-}(-i\omega_{0})}{Q^{-}(0)}.\]
The condition for a bond to snap is that the total length of the bond
reach $\lspace\lambda_{f}$. Note from the definition in Eq. \prettyref{eq:TL12}
that $U$ gives only half the bond length. Therefore\begin{eqnarray*}
\lambda_{f}^{2} & = & \frac{1}{4}\lambda_{x}^{2}+\frac{(2U_{0})^{2}}{\lspace^{2}}\\
\Rightarrow2U_{0} & = & \lspace\sqrt{(\lambda_{f}^{2}-\lambda_{x}^{2}/4)}=\frac{\sqrt{3}}{2}\lspace\lambda_{y}\frac{Q^{-}(-i\omega_{0})}{Q^{-}(0)}\\
\noalign{\vskip-20pt}\end{eqnarray*}
\[
\Rightarrow\frac{\lambda_{y}}{\sqrt{(4\lambda_{f}^{2}-\lambda_{x}^{2})/3}}=\frac{Q^{-}(0)}{Q^{-}(-i\omega_{0})}.\]

Employing \prettyref{eq:L93a.2;b} to obtain an explicit representation
of $Q^{-}$ leads to Eq. \prettyref{eq:result}.

\bibliographystyle{/usr/share/texmf/bibtex/bst/natbib/elsart-harv}
\bibliography{/home/marder/grants/marder,/home/marder/references/rubber_acoustic_attenuation,/home/marder/references/fracture}

\end{document}